\documentclass[camera,letterpaper,nomarginnotes,nonarrowgutter]{jpaper}

\usepackage{amsmath,amssymb,amsfonts}
\usepackage{graphicx}
\usepackage{textcomp}
\usepackage{xcolor}
\usepackage{fancyhdr}
\usepackage[export]{adjustbox}
\usepackage{algorithm}
\usepackage{algorithmicx}
\usepackage{algpseudocode}

\usepackage{soul}
\usepackage{arevmath}
\usepackage{setspace}
\usepackage{xspace}
\usepackage[utf8]{inputenc}
\usepackage{blindtext, subfig}
\usepackage{fancyhdr}
\usepackage[colorlinks=true, linkcolor=black, anchorcolor=black, citecolor=black, filecolor=black, menucolor=black, runcolor=black, urlcolor=black]{hyperref}
\usepackage[normalem]{ulem}
\usepackage{datetime}

\usepackage[italic]{mathastext}
\usepackage{libertine}
\usepackage[T1]{fontenc}
\usepackage{textcomp}
\usepackage[varqu,varl]{zi4}
\usepackage[all]{nowidow}
\usepackage{flushend}
\usepackage{multirow}
\usepackage{paralist}
\usepackage{soul}

\usepackage{balance}

\usepackage{circledsteps}

\usepackage[compress,sort]{cite}

\newcommand{\graph}[0]{\emph{ALPHA-PIM}}

\begin{document}
\bstctlcite{IEEEexample:BSTcontrol}
\title{\huge \graph:\\ Analysis of Linear Algebraic Processing for High-Performance Graph Applications on \\ a Real Processing-In-Memory System
}

\newcommand{\affilsfu}[0]{\textsuperscript{\S}}
\newcommand{\affilETH}[0]{\textsuperscript{$\dagger$}}
\newcommand{\affiltoronto}[0]{\textsuperscript{$\ddagger$}}
\newcommand{\affilnvidia}[0]{\textsuperscript{$\nabla$}}
\newcommand{\affilbeirut}[0]{\textsuperscript{$\ast$}}

\author{
{Marzieh~Barkhordar\affilsfu}~~~%
{Alireza~Tabatabaeian\affilsfu}~~~%
{Mohammad~Sadrosadati\affilETH}~~~%
{Christina~Giannoula\affiltoronto}\\%
{Juan~Gomez-Luna\affilnvidia}~~~%
{Izzat~El-Hajj\affilbeirut}~~~%
{Onur~Mutlu\affilETH}~~~%
{Alaa~R. Alameldeen\affilsfu}
\vspace{2mm}\\%
\emph{{\affilsfu Simon Fraser University~~~~~ 
\affilETH ETH Z{\"u}rich~~~~~}}\\ \emph{{\affiltoronto  University of Toronto 
~~~~~\affilnvidia NVIDIA
~~~~~\affilbeirut American University of Beirut}
}%
}

\maketitle
\thispagestyle{plain}

\begin{abstract}
Processing large-scale graph datasets is computationally intensive and time-consuming.
Processor-centric CPU and GPU architectures, commonly used for graph applications, often face bottlenecks caused by extensive data movement between the processor and memory units due to low data reuse. As a result, these applications are often memory-bound, limiting both performance and energy efficiency due to excessive data transfers. 
Processing-In-Memory (PIM) offers a promising approach to mitigate data movement bottlenecks by integrating computation directly within or near memory. Although several previous studies have introduced custom PIM proposals for graph processing, they do not leverage real-world PIM systems. 


This work aims to explore the capabilities and characteristics of common graph algorithms on a real-world PIM system to accelerate data-intensive graph workloads. To this end, we (1) implement representative graph algorithms on UPMEM’s general-purpose PIM architecture; (2) characterize their performance and identify key bottlenecks; (3) compare results against CPU and GPU baselines; and (4) derive insights to guide future PIM hardware design.

Our study underscores the importance of selecting optimal data partitioning strategies across PIM cores to maximize performance.
Additionally, we identify critical hardware limitations in current PIM architectures and emphasize the need for future enhancements across computation, memory, and communication subsystems. Key opportunities for improvement include increasing instruction-level parallelism, developing improved DMA engines with non-blocking capabilities, and enabling direct interconnection networks among PIM cores to reduce data transfer overheads.



\end{abstract}


\section{Introduction}\label{sec:introduction}
Graph data structures are widely used to model complex problems across domains due to their simplicity and versatility \cite{intro-graphapp-social-polites2008centrality,intro-graphapp-social-azad2015parallel,intro-graphapp-chemical-ivanciuc1999graph, intro-graphapp-NLP-manning2014stanford,intro-graphapp-NLP-hahn1984computing,intro-graphapp-NLP-miller1995wordnet}. As real-world graphs scale to billions of nodes and edges, parallel computing systems and algorithms have been developed to improve performance. However, efficient parallel graph processing remains challenging due to irregular memory access, load imbalance, and low arithmetic intensity \cite{background-linearAlgebraicGraph, motivation-bulk-anavaram}.

Linear algebra is a 
promising paradigm
for modeling graph applications using fundamental matrix operations \cite{background-linearAlgebraicGraph}. 
It offers a concise set of primitives that, combined with algebraic semirings, efficiently represent diverse graph algorithms \cite{background-linearAlgebraicGraph,background-semiring1,background-semiring2,background-semiring3,background-semiring4,background-semiring5,background-semiring6}. Graph algorithms are often more compact and easier to express using sparse-matrix linear algebra. Describing an algorithm in this format 
eliminates the need for additional data structures and can be directly implemented in array-based programming environments, making it easier to optimize. 


Given the sparsity of adjacency matrices in graph datasets, matrices are stored in a compressed format. Consequently, sparse matrix operations, such as sparse matrix-vector multiplication (SpMV) are used to implement graph applications. However, these operations involve indirect memory references due to the compressed storage format and 
irregular memory accesses to the input vector
caused by sparsity, which hinders their ability to achieve peak performance in processor-centric CPU and GPU architectures. Many real-world graph datasets across various application domains are highly sparse, making SpMV a memory-bound kernel in processor-centric systems that is limited by data movement between memory and processors \cite{memory-bound-benchmarking, memory-bound-benchmarking2, memory-bound-conflict,memory-bound-multicore,memory-bound-sparse,memory-bound-sparsity, accelerator-pim-scalable,ipc-memory-bound-comparison,ipc-memory-bound-energy}.

Processing-In-Memory (PIM) offers a promising solution to the data movement bottleneck by moving computation closer to the data, and integrating processing capabilities within memory chips \cite{motivaiton-pim-co-ml, motivation-pim-concurrent, motivation-pim-nda, motivation-pim-continuous, motivation-pim-google, motivation-pim-neural, motivation-pim-processing, PyGimSIGMETRICS25, SynCronHPCA2021}. Several prior PIM proposals aim to accelerate graph processing \cite{accelerator-pim-scalable,accelerator-pim-graphp,accelerator-pim-graphh,accelerator-pim-graphpim, accelerator-pim-sisa, accelerator-pim-pim-enabled, accelerator-pim-conda, accelerator-pim-lazy, accelerator-pim-graphq, PyGimSIGMETRICS25}. However, none provides a comprehensive evaluation of general-purpose real-world PIM systems. To our knowledge, this is the first paper to comprehensively study the opportunities and limitations of linear algebraic graph processing using a real PIM system. 

We analyze 
the execution time breakdown of SpMV implementations from SparseP \cite{pim-sparsep}, the first open-source SpMV library specifically designed for real PIM systems. Prior work indicates that traversal-based graph applications often require multiple iterations of matrix-vector multiplication, with highly-sparse input vectors in several iterations \cite{related-framework-graphblast, motivaion-spmspv-parallel, motivation-spmspv-fast, motivation-spmspv-owen}. We observe that SpMV kernels suffer from high data transfer costs due to using a dense input vector format. 
To address this issue, we  further minimize these data transfer costs through sparse-matrix sparse-vector multiplication (SpMSpV) \cite{spmspv-parallel, spmspv-tile}. Using a compressed format for the input vector in SpMSpV reduces the significant overhead associated with loading input vectors into DRAM banks.

Our goal is to understand the capabilities and characteristics of a real-world PIM system for accelerating graph applications.
To do so, we introduce \graph, the first 
linear algebraic graph application framework designed for a real-world PIM architecture. 
We choose UPMEM’s PIM system \cite{background-upmem} for our study because it is the first commercially available general-purpose PIM architecture, combining conventional 2D DRAM arrays with general-purpose processing cores, known as DRAM Processing Units (DPUs), on the same chip.
We design and evaluate \graph~in five key steps. First, we implement and evaluate different versions of the SpMSpV kernel using various compressed matrix formats (CSR, CSC, COO) and partitioning strategies (row-wise, column-wise, 2D) to identify the most efficient approach for graph applications. 
Second, we implement three well-known linear algebraic graph applications: Breadth-First Search (BFS) \cite{app-bfs-improving,app-bfs-local}, Single Source Shortest Path (SSSP) \cite{app-sssp-fast,app-sssp-highway,app-sssp-hub}, and Personalized PageRank (PPR) \cite{app-ppr}, using both SpMV and SpMSpV kernels. Third, we characterize performance of these graph applications and identify key bottlenecks. Fourth, we compare our graph implementation against state-of-the-art graph processing frameworks on both CPU and GPU platforms. Finally, we explore implications for future PIM hardware enhancements, emphasizing the importance of algorithm–hardware co-design to reduce communication costs and improve the performance of linear-algebraic graph processing on a real-world PIM system.



Our results highlight two major observations. First, it is critically important to select appropriate partitioning strategies and compressed matrix formats to maximize performance on PIM systems.
In our experiments, we observed up to a 25$\times$ difference in SpMSpV's execution time between the best and worst-performing strategies, underscoring the importance of careful strategy and format selection for maximizing performance.
Second, the performance of these applications is mainly limited due to (i) computation-side bottlenecks, e.g., 
idle DPU 
cycles caused by structural hazards in the revolver pipeline; (ii) memory-side bottlenecks, e.g., idle DPU cycles due to waiting on memory operations; and (iii) communication-side bottlenecks from loading and retrieving vectors between iterations, compounded by the lack of inter-DPU communication. 
Based on these observations, we suggest several hardware optimizations to enhance the performance of graph processing on a real PIM architecture. 

\vspace{4pt}
In this paper, we make the following contributions:
\begin{compactitem}
    \vspace{4pt}
    \item We present the first linear-algebraic implementation of several graph applications 
     on a real-world PIM system (UPMEM), and conduct a thorough design-space exploration of data structures used to represent graph datasets and input vectors.
    \vspace{4pt}
    \item We implement and conduct the first comprehensive study of the widely used SpMSpV kernel for graph applications on 
    a state-of-the-art real-world PIM system.
    \vspace{4pt}
    \item We compare the performance of graph implementations on the state-of-the-art UPMEM PIM system vs. conventional CPU and GPU systems. We show that \graph~achieves kernel speedups of 10.2$\times$ / 48.8$\times$  / 3.6$\times$  and total execution speedups of 2.6$\times$  / 10.4$\times$  / 1.7$\times$  for BFS / SSSP / PPR, respectively, over a conventional CPU baseline, while also delivering better compute utilization than both CPU and GPU systems.

\end{compactitem}


\section{Background and Motivation} 
\label{sec:background}

\subsection{Graph Applications Using Linear Algebra}
Graph algorithms can be expressed as linear algebra operations due to the duality between a graph and its adjacency matrix~\cite{background-linearAlgebraicGraph}. For a graph $G = (V, E)$ with $N$ vertices, its $N \times N$ adjacency matrix $A$ has $A_{ij} = 1$ if $v_i$ is adjacent to $v_j$, and 0 otherwise. This enables algorithms like BFS to be implemented as iterative matrix-vector multiplications, e.g., $v = A^T v$, where non-zero elements in $v$ indicate visited nodes. To support a broader range of algorithms, this framework can be extended using a \emph{semiring}~\cite{background-linearAlgebraicGraph,background-semiring1,background-semiring2,background-semiring3,background-semiring4,background-semiring5,background-semiring6}, which generalizes addition and multiplication to $\oplus$ and $\otimes$ with algebraic properties. For instance, SSSP can be expressed using the $(\min, +)$ semiring over extended real numbers.

Real-world graphs are typically sparse, making efficient matrix storage essential. Three widely used compressed formats are Coordinate List (COO), Compressed Sparse Row (CSR), and Compressed Sparse Column (CSC)~\cite{background-coo,background-coo-csr,background-csr,background-linearAlgebraicGraph}. 
COO stores non-zero elements as (i, j, value) tuples; while simple and parallel-friendly, it lacks row-wise grouping, leading to scattered updates that require atomics and reduce efficiency for large matrices.
CSR compresses row indices using three arrays: $values$, $col\_indices$, and $row\_ptr$, supporting efficient row-wise operations. Conversely, CSC compresses columns via $values$, $row\_indices$, and $col\_ptr$, making it more suitable for column-wise operations like vector-matrix multiplications.

\subsection{Linear Algebraic Graph Applications in Processor-Centric Systems}

Linear algebra provides an effective framework for modeling graph applications via matrix operations, simplifying parallel algorithm implementation~\cite{background-linearAlgebraicGraph}. By combining linear-algebraic primitives with algebraic semirings, diverse graph algorithms can be efficiently expressed. For example, in BFS, replacing addition/multiplication with OR/AND reduces computation and allows early termination, improving runtime. These primitives also enable 2D matrix partitioning—a critical capability for scaling to large graphs that many other frameworks lack~\cite{related-framework-graphblast}. As a result, several linear-algebraic graph frameworks have been developed for CPUs and GPUs to leverage these advantages~\cite{related-framework-graphblast, accelerator-algebra-slimsell, frameworks-highlevel-gbtl, app-nvgraph}.


Despite their advantages, linear-algebraic graph frameworks on CPUs and GPUs face performance challenges due to the sparsity of real-world graphs, which necessitates compressed formats and sparse matrix operations like SpMV and SpMM. Studies show these operations often fall short of peak performance~\cite{memory-bound-benchmarking, memory-bound-benchmarking2, memory-bound-conflict,memory-bound-multicore,memory-bound-sparse,memory-bound-sparsity, accelerator-pim-scalable,ipc-memory-bound-comparison,ipc-memory-bound-energy}, primarily due to algorithmic characteristics, storage formats, and sparsity patterns. SpMV is especially memory-bound: its irregular access pattern hinders cache locality; each matrix element is accessed only once, i.e., no temporal locality; and auxiliary structures for non-zero indexing add bandwidth pressure and increase contention.

In contrast, PIM systems are well suited to address the challenges of linear-algebraic graph processing. By moving computation closer to memory and reducing data movement, PIM systems can exploit high levels of parallelism and provide substantial memory bandwidth \cite{motivaiton-pim-co-ml, motivation-pim-concurrent, motivation-pim-nda, motivation-pim-continuous, motivation-pim-google, motivation-pim-neural, motivation-pim-processing}, which significantly improve performance of memory-bound operations like SpMV.
Several prior PIM proposals have aimed to accelerate graph processing \cite{accelerator-pim-scalable,accelerator-pim-graphp,accelerator-pim-graphh,accelerator-pim-graphpim, accelerator-pim-sisa, accelerator-pim-pim-enabled, accelerator-pim-conda, accelerator-pim-lazy, accelerator-pim-graphq}. However, none have comprehensively evaluated real general-purpose PIM architectures. This motivates our exploration of the capabilities and characteristics of popular graph algorithms on UPMEM’s PIM architecture. 
\label{sec:motivation}

\subsection{The UPMEM PIM Architecture}


\subsubsection{\textbf{System Organization}}
\begin{figure}[!ht]
  \centering
  \includegraphics[width=1\linewidth]{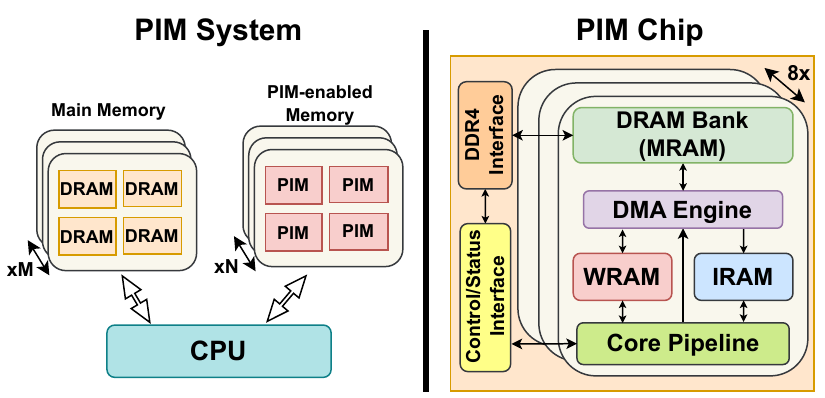}
  \caption{High-level organization of the UPMEM PIM system with a host CPU, DRAM main memory, and PIM-enabled memory (left), and a more detailed view of the UPMEM PIM chip (right) \cite{background-upmem}.}
\label{fig:background-upmem}
\end{figure}

Figure~\ref{fig:background-upmem} illustrates the UPMEM architecture \cite{background-upmem,background-upmem-website}. A UPMEM system consists of a host CPU (e.g., x86, ARM64, or 64-bit RISC-V), standard DRAM modules, and PIM-enabled memory modules designed in the DDR4-2400 DIMM form factor. These modules house multiple PIM cores, called DRAM Processing Units (DPUs), with each module comprising multiple ranks, each containing 8 DPUs. A typical configuration supports up to 2,560 DPUs across 20 double-ranked DIMMs. UPMEM DIMMs connect to the host CPU via memory channels and integrate PIM chips, each pairing a DPU core with a DRAM bank.
The host CPU manages data transfers between its memory and the DPUs' Main RAM (MRAM) using a transposition library provided by the UPMEM Software Development Kit (SDK). This library ensures that data layout transformations are handled transparently, facilitating efficient data movement between main memory and PIM-enabled memory.
The UPMEM SDK supports both serial and parallel data transfers. Parallel transfers can be performed across multiple MRAM banks which 
allows for concurrent CPU-DPU and DPU-CPU data transfers.

\subsubsection{\textbf{DPU Architecture}}\label{subsec:dpu-architech}

Each DPU is a multithreaded, in-order 32-bit RISC core designed with a 14-stage pipeline. It features a 64 MB DRAM bank (MRAM), a 64 KB scratchpad memory (WRAM), and a 24 KB instruction memory (IRAM). The MRAM serves as the main memory, the WRAM as a high-speed temporary storage, and the IRAM stores the instructions to be executed. The DPU can concurrently run up to 24 hardware PIM threads, i.e., tasklets, which share access to the WRAM, IRAM, and MRAM. 
The DPU's unique "revolver pipeline" design enforces a scheduling constraint where consecutive instructions within the same thread must be dispatched 11 cycles apart. This constraint simplifies the microarchitecture by eliminating the need for complex data forwarding and pipeline interlocks. The register file within the DPU is split into even and odd sections, preventing simultaneous access to multiple registers within the same group due to structural hazards.

\subsubsection{\textbf{DPU Programming}}
Programming the UPMEM PIM system follows the single-program multiple-data (SPMD) model, where all tasklets execute the same program across multiple DPUs on distinct data partitions. Efficient execution requires careful input data partitioning to ensure workload balance. On the host side, the CPU manages DPU allocation, loads program binaries and data, and controls execution. Programs are written in UPMEM’s C-like language and compiled using an LLVM-based toolchain to produce binaries for both host and DPUs. On the DPU side, tasklets access data via WRAM, requiring explicit transfers from MRAM using DMA instructions. Since computation is limited to WRAM-resident data, careful data movement is critical. Intra-DPU synchronization uses WRAM or MRAM primitives like mutexes and barriers, while inter-DPU coordination must go through the host CPU due to the absence of direct DPU-to-DPU communication.

\section{Limitations of SpMV Implementation for Graph Applications on UPMEM}
\label{sec:limitation-spmv}

Linear-algebraic graph applications often use SpMV to process sparse matrices efficiently by avoiding redundant computations on zero elements. SpMV multiplies a sparse matrix of size $M \times N$ by a dense vector of size $N \times 1$, producing an output vector of size $M \times 1$. SparseP \cite{pim-sparsep}, the first SpMV library for real PIM systems, evaluates SpMV using 1D and 2D partitioning. In 1D, the matrix is row-partitioned and each DPU receives the full input vector, enabling local computation with minimal merging. In 2D, matrices are tiled, transferring only relevant input vector segments, but overlapping tiles increase CPU merge overhead. SparseP’s top-performing variants are COO.nnz (1D) and DCOO (2D), both using static, equal-sized COO-format tiles.

\begin{figure}[!ht]
  \centering
  \includegraphics[width=1\linewidth]{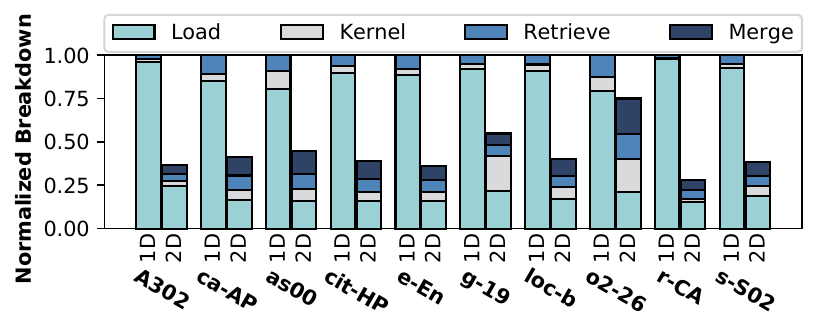}
  \caption{Execution time breakdown using 2048 DPUs and int32 data type for 1D/2D SpMV partitioning (normalized to 1D).}
\label{fig:linalg-spmv}
\end{figure}

In Figure~\ref{fig:linalg-spmv}, we report the execution time breakdown of two top 1D/2D partitioning methods for SpMV in SparseP using 2048 DPUs and INT32 data (normalized to 1D partitioning). The total time is divided into four parts: (i) loading the input vector from the CPU to DPU DRAM (Load); (ii) kernel execution on DPUs (Kernel); (iii) retrieving results from DPUs to the CPU (Retrieve); and (iv) merging partial results on the CPU (Merge). Section \ref{sec:evaluation-methodology} presents methodology details. 
We note that
1D partitioning incurs a high cost for transferring and broadcasting the input vector to each DRAM bank in the PIM core. While 2D partitioning reduces the transfer cost, it adds extra data transfer overhead for gathering and merging results from PIM memory to the host CPU.

SparseP addresses the high cost of input vector loads in SpMV using 2D partitioning, but its reliance on the dense format limits its effectiveness in graph applications.
Previous works \cite{related-framework-graphblast, motivaion-spmspv-parallel, motivation-spmspv-fast, motivation-spmspv-owen} show that in many iterations of linear-algebraic graph applications, input vectors are highly sparse—e.g., in BFS, only a small subset of vertices (the frontier) is active at each level, leaving most entries as zero.
To quantify the input vector sparsity in BFS, 
we measured its \emph{density} (the ratio of non-zero elements to the total number of nodes, expressed as a percentage) across several datasets. Our results indicate that, for most cases, the input vector's density remains below 50\% during the first half of the iterations. 
Therefore, there is still potential to reduce the cost of input vector loads in linear-algebraic graph applications
by exploring the potential of Sparse Matrix-Sparse Vector (SpMSpV) multiplication \cite{spmspv-parallel, spmspv-tile}.




\section{\graph~Design and Optimizations}
\label{sec:graph-pim}
In this paper, we target graph applications with low arithmetic intensity and limited data reuse, such as BFS, SSSP, and PPR. These traversal-based algorithms rely on multiple iterations of matrix-vector multiplication, making them memory-bound and ideal candidates for PIM systems.
These algorithms require the use of both SpMV and SpMSpV due to the varying sparsity of the input vector across iterations \cite{frameworks-single-direction, related-framework-graphblast}. 
While SparseP \cite{pim-sparsep} provides a comprehensive analysis of SpMV, we focused on developing an efficient SpMSpV implementation for UPMEM.

Our approach involves three key steps: \textcircled{1} We create and evaluate various SpMSpV implementations using different compressed matrix formats and partitioning strategies to identify the most efficient option for graph applications; \textcircled{2} we employ an empirical cost model to determine the optimal kernel (SpMV or SpMSpV) based on the input vector’s sparsity and graph dataset characteristics; and \textcircled{3} we extend these implementations to support a range of traversal-based graph applications by incorporating different semirings into both SpMV and SpMSpV kernels.

\subsection{SpMSpV Implementation on UPMEM}
The SpMSpV operation, expressed as $y \leftarrow Ax$, multiplies a sparse matrix $A$ by a sparse vector $x$ to produce a vector $y$ \cite{spmspv-parallel, spmspv-tile}. Both the input matrix $A$ and the input vector $x$ are stored in compressed formats to efficiently manage their sparsity. 
Focusing on the non-zero elements of $x$ reduces the number of computations and memory accesses, making SpMSpV suitable for large-scale, sparse datasets, unlike SpMV which assumes a dense input vector.

In this work, we implement SpMSpV using three compressed matrix formats: COO, CSR, and CSC. In SpMSpV using CSR or COO formats, the computation involves iterating through the rows of the input matrix, with the input vector in a compressed format, ensuring that only non-zero elements are processed. This requires considering the entire adjacency matrix, and matching each row element with the corresponding non-zero vector elements based on their indices. In contrast, SpMSpV using the CSC format focuses on active columns of the matrix, meaning it processes only those columns whose indices match the indices of non-zero elements in the input vector. This approach reduces the overall number of operations by excluding columns that do not align with non-zero entries in the input vector, thereby improving efficiency.


To execute SpMSpV on UPMEM, we perform four steps: \textcircled1 Load the input vector into the UPMEM's DRAM banks (Load); \textcircled2 run the SpMSpV kernel on DPUs (Kernel); \textcircled3 transfer results from DRAM banks to the host CPU (Retrieve); and \textcircled4 merge partial results to form the final output vector on the host CPU (Merge). 
We exclude the time needed for loading the matrix into UPMEM memory from our analysis, as this step is typically amortized over multiple kernel iterations or overlapped with other computations, making it negligible in real-world graph applications. We next describe SpMSpV implementation details including partitioning, parallelism, and memory access patterns. 

\subsubsection{\textbf{Partitioning Across DPUs}}

\begin{figure*}[!ht]
  \centering
  \includegraphics[width=0.9\linewidth]{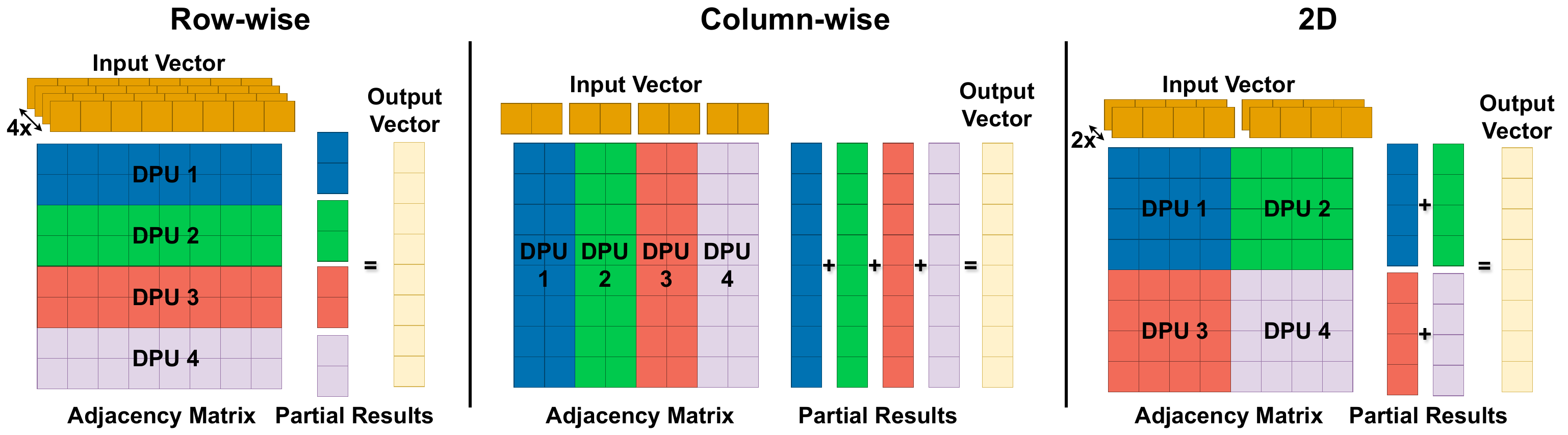}
  \caption{Partitioning Strategies of Adjacency Matrix.}
\label{fig:partitioning}
\end{figure*}

Partitioning strategies in UPMEM face different challenges than CPU/GPU due to architectural constraints. Unlike CPU/GPU, which optimize memory access via caching and coalescing, DPUs have limited MRAM and WRAM, restricting the amount of data each DPU can store and process efficiently. UPMEM lacks direct inter-DPU communication, which increases communication overhead due to relying on the host CPU for communication across DPUs. Efficient partitioning must minimize CPU-DPU transfers to reduce data movement and performance loss. According to Figure~\ref{fig:partitioning}, we partition the matrix in three ways: Row-wise, Column-wise, and 2D. 
Each method is tailored to different aspects of parallel computation and data access patterns. We next discuss each of these partitioning strategies in more detail.

\vspace{4pt}
\noindent\textbf{Row-wise Partitioning.} 
This involves dividing the adjacency matrix into 
$D$ partitions, each corresponding to a set of rows, where $D$ is the number of DPUs utilized in the computation.
Each partition is then converted into the desired graph representation format (COO, CSR, or CSC) and loaded into the MRAM of a DPU. This method requires copying the entire compressed input vector into the DRAM bank of each PIM core. 
Each DPU computes a portion of the output vector result, but the computation is carried out exclusively by the PIM cores, eliminating the need for a merging step. We implemented three row-wise partitioning versions for SpMSpV, utilizing three different compressed formats: CSR, COO, and CSC (referred to as CSC-R). 

\vspace{4pt}
\noindent\textbf{Column-wise Partitioning.} This
involves dividing the adjacency matrix into $D$ partitions by column, where $D$ is the number of DPUs utilized in the computation. Each partition is converted into the desired graph representation. Only a subset of compressed input vector elements is loaded into each PIM core's DRAM bank. Each PIM core generates a complete output vector of size equal to the number of nodes, representing a partial result of the final output vector, which then requires merging on the host CPU. For column-wise partitioning, we only utilize the CSC compressed format, as the other formats (CSR and COO) are 
inefficient for this partitioning. We refer to this version as CSC-C for brevity. 

\vspace{4pt}
\noindent\textbf{2D Partitioning.} 
We divide the adjacency matrix into a grid of tiles
matching the number of available 
DPUs. This method seeks to balance the trade-off between input vector transfer and output vector transfer
as each DPU handles only a portion of the input and output vector stored in its DRAM bank. However, when DPUs are assigned overlapping tiles—where tiles share rows of the matrix that are divided among multiple tiles—each DPU generates numerous partial results for the output vector. These results are then transmitted to the host CPU, where they are merged to form the final output vector. In our work, the CPU cores execute the merge step in parallel using the OpenMP API. 
Similar to column-wise partitioning, we exclusively use the CSC compressed format for 2D partitioning (denoted as CSC-2D) due to the inefficiency of the other formats. 


\subsubsection{\textbf{Thread-Level Parallelism}}
To leverage the high memory bandwidth of PIM banked DRAM, each PIM core executes multiple hardware threads. We assign one partition to each core and ensure balanced computation across its threads by evenly distributing the workload, either by row or by column 
depending on the matrix compression format. This thread-level workload balancing helps minimize kernel execution time.

\subsubsection{\textbf{Memory Access Patterns and Kernel Logic}}
We briefly describe how threads access data from and to their local DRAM bank during kernel execution. 

First, SpMSpV uses streaming accesses for reading non-zero values and their indices. To exploit spatial locality and high bandwidth, each thread fetches large chunks of bytes in a coarse-grained fashion from DRAM into WRAM (UPMEM’s scratchpad), then accesses them in a fine-grained fashion. This also applies to loading the compressed input vector in SpMSpV, where memory access patterns are more localized than in SpMV. In SpMV, input vector accesses are input-driven, determined by the column indices of the non-zero matrix elements, leading to irregular memory accesses. 

Second, threads buffer partial results for the same output row in WRAM, exploiting temporal locality before writing back to DRAM. Since multiple threads may update the same output vector elements concurrently, synchronization primitives (mutexes and barriers) are used to ensure correctness and avoid race conditions.

\subsection{Adaptive SpMSpV–SpMV Switching}

Traversal-based graph algorithms such as BFS, SSSP, and PPR rely on iterative matrix-vector multiplications. A critical characteristic of these algorithms is that the sparsity of the input vector changes across iterations—typically starting sparse and becoming denser as the algorithm progresses. 
Since SpMSpV is more efficient when the input vector is sparse while SpMV performs better when it is dense, switching between them at the right point can significantly reduce total execution time. 

\begin{figure}[!ht]
  \centering
  \includegraphics[width=1\linewidth]{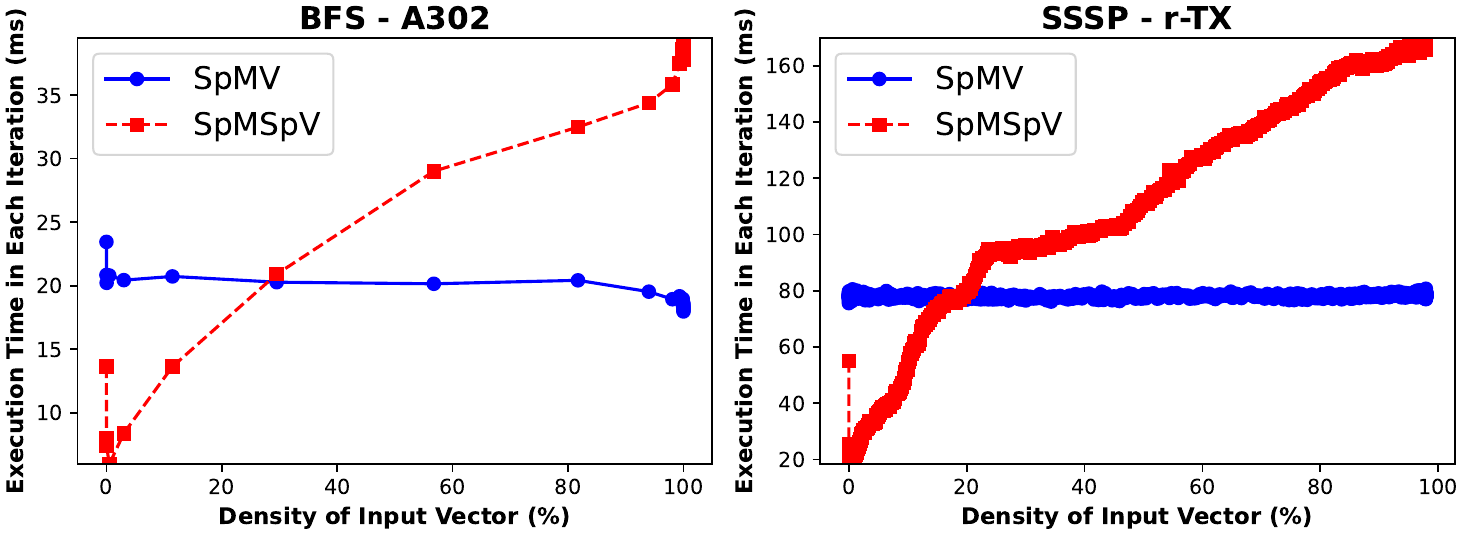}
    \caption{Execution time per iteration on UPMEM for BFS and SSSP of two datasets using SpMV and SpMSpV.}
\label{fig:graph-iterations}
\end{figure}

We analyzed various SpMV and SpMSpV implementations to identify the most efficient options for graph applications, selecting CSC-2D for SpMSpV and DCOO (the best 2D-partitioned SpMV from SparseP) for SpMV (see Sections~\ref{sec:spmspv-analysis} and~\ref{sec:limitation-spmv} for details). Figure~\ref{fig:graph-iterations} compares per-iteration total execution time for two algorithms and datasets using two strategies: \textcircled{1} SpMV-only for all iterations and \textcircled{2} SpMSpV-only for all iterations. The x-axis shows iteration time (ms), and the y-axis shows input vector density.
We observe that SpMSpV performance scales with input sparsity, while SpMV remains steady across iterations.
This experiment shows how performance shifts between these kernels at different densities.

\subsubsection{\textbf{Empirical Cost Model and Switch Point}}
We define the optimal switching point as the input vector density—the ratio of non-zero entries to the total number of nodes, expressed as a percentage—at which SpMV begins to outperform SpMSpV. This threshold varies based on the structural characteristics of the graph dataset. 

To identify these switching behaviors, we analyzed a diverse set of real-world graphs and found a consistent trend: \textcircled{1} Regular graphs (e.g., road networks), which have low average degrees and uniform degree distributions, exhibit an optimal switching point around 20\% density, and \textcircled{2} Scale-free graphs (e.g., web and social networks), which have skewed degree distributions and higher average degrees, exhibit a switching point around 50\% density. Based on this observation, we categorize datasets into two classes—regular and scale-free—and use this classification to guide the kernel selection process.

To automate this selection at runtime, we design a lightweight decision tree model trained on a diverse set of real-world graphs. Our decision tree distinguishes between two dominant graph types—regular and scale-free—which cover most real-world graphs, enabling a practical and generalizable kernel switching mechanism. The model takes as input two graph features: the average node degree and the standard deviation of degrees. It classifies the graph into one of the two categories and selects the appropriate switching threshold. At runtime, we monitor the input vector density in each iteration. Once the density exceeds the threshold predicted by the decision tree, we transition from the SpMSpV kernel to the SpMV kernel. 

We conducted a sensitivity analysis on the switching threshold and found that a 10\% deviation results in less than 5\% increase in total runtime on average. For example, as shown in Figure \ref{fig:graph-iterations}, using a 60\% switching threshold instead of 50\% for the 'A302' dataset increases the total runtime by only 2.5\%.\footnote{We note that the 'A302' has an iteration at around 30\% input vector density followed by an iteration with around 60\% density, so no single iteration has a 50\% density. Our switching point is the first iteration after exceeding 50\% density.} This demonstrates that our adaptive model is robust to modest threshold errors.    

\subsubsection{\textbf{Runtime Overhead and Practicality}} Both the required graph statistics and the classification are computed once during pre-processing on the CPU. The model is computationally lightweight, incurs negligible runtime overhead, and runs efficiently on the CPU. This strategy enables adaptive and dataset-aware kernel selection, ensuring that the most efficient matrix-vector multiplication kernel is used throughout execution.

\section{Evaluation Methodology} \label{sec:evaluation-methodology}


\subsection{Graph Applications}
We implement three widely-used graph algorithms—BFS, SSSP, and PPR—each with real-world applications \cite{background-semiring2,app-basic-graph,app-basic-graphnetwork}. BFS finds neighboring nodes \cite{app-bfs-improving,app-bfs-local}; SSSP enables routing in road networks \cite{app-sssp-fast,app-sssp-highway,app-sssp-hub}; PageRank ranks web pages \cite{app-pr-pagerank}, and its personalized variant, PPR, emphasizes node importance from a specific source for recommendations and local search \cite{app-ppr}. All rely on matrix-vector multiplication with different semirings (see Table~\ref{tab:algorithm-semirings}). A broader set is listed in \cite{background-linearAlgebraicGraph}. These algorithms are representative and our findings generalize to similar workloads.


\begin{table}[hbt!]
\begin{center}
\caption{Algorithms with their semirings.
\strut}
\label{tab:algorithm-semirings}
\begin{tabular}{|l|c|c|}
\hline
\textbf{Algorithm} & \textbf{Semiring} & \textbf{Operations ($\oplus$ and $\otimes$)} \\ \hline\hline
BFS                & ${0, 1}$          & $|~\&$               \\ \hline
SSSP               & $R \cup {\infty}$ & $min~+$             \\ \hline
PPR                 & $R$               & $+~\times$          \\ \hline
\end{tabular}
\end{center}
\end{table}

\subsection{Tools}\label{sec:tools}
For the real machine experiments, we conduct our evaluation on an UPMEM PIM system featuring a 2-socket Intel Xeon Silver 4110 CPU \cite{tools-cpu} running at 2.10 GHz (host CPU), 128 GB of standard DDR4-2400 main memory \cite{tools-ddr4}, and 20 UPMEM PIM DIMMs offering 160 GB of PIM-capable memory and 2560 DPUs. We also use PIMulator, an advanced UPMEM simulator \cite{tools-upimulator} to gain deeper insights and gather additional performance metrics.

\subsection{Data Sets}
We selected 65 graph datasets from GraphChallenge \cite{datasets-graph}, spanning diverse domains. Table~\ref{tab:dataset-charc} summarizes 13 representative datasets used in our experiments, showing their names, abbreviations, edge/node counts, average degrees, degree standard deviations, and sparsity. Sparsity is defined as $NNZ/ N^2$, where $NNZ$ is the number of edges and $N$ is the number of graph nodes. These cover several of our evaluations. Details on the full set are available in \cite{datasets-graph-snap}.

\begin{table}[h]\centering
\caption{The characteristics of the example 13 real-world datasets}
\label{tab:dataset-charc}
\resizebox{\columnwidth}{!}{%
\begin{tabular}{|l|c|c|c|c|c|c|}
\hline
\textbf{Datasets} 
& \textbf{Abbreviation} & \textbf{Edge} & \textbf{Node} & \textbf{AVG-Deg} & \textbf{Deg-std} & \textbf{Sparsity} \\ \hline\hline
amazon0302 & A302 & 899792 & 262111 & 6.86 & 5.41 & 1.31E-05 \\
\hline
as20000102 & as00 & 12572 & 6474 & 3.88 & 24.99 & 3.00E-04 \\
\hline
ca-GrQc & ca-Q & 14484 & 5242 & 5.52 & 7.91 & 5.27E-04 \\
\hline
cit-HepPh & cit-HP & 420877 & 34546 & 24.36 & 30.87 & 3.53E-04 \\
\hline
email-Enron & e-En & 183831 & 36692 & 10.02 & 36.1 & 1.37E-04 \\
\hline
facebook\_combined & face & 88234 & 4039 & 43.69 & 52.41 & 5.41E-03\\
\hline
graph500-scale18 & g-18 & 3800348 & 174147 & 43.64 & 229.92 & 1.25E-04\\
\hline
loc-brightkite\_edges & loc-b & 214078 & 58228 & 7.35 & 20.35 & 6.31E-05\\
\hline
p2p-Gnutella24  & p2p-24 & 65369 & 26518 & 4.93 & 5.91 & 9.30E-05 \\
\hline
roadNet-TX  & r-TX & 1541898 & 1088092 & 2.78 & 1.0 & 1.01E-06\\
\hline
soc-Slashdot0902  & s-S02 & 504230 & 82168 & 12.27 & 41.07 & 7.47E-05\\
\hline
soc-Slashdot0811  & s-S11 & 469180 & 77360 & 12.12 & 40.45 & 7.84E-05\\
\hline
flickrEdges  & flk-E & 2316948 & 105938 & 43.74 & 115.58 & 2.06E-04\\
\hline
\end{tabular}
}
\end{table}







\section{Performance Evaluation}\label{sec:results}

\subsection{Trade-Offs in SpMSpV Strategies}\label{sec:spmspv-analysis}

\begin{figure*}[h!]
  \centering
  \includegraphics[width=1\linewidth]{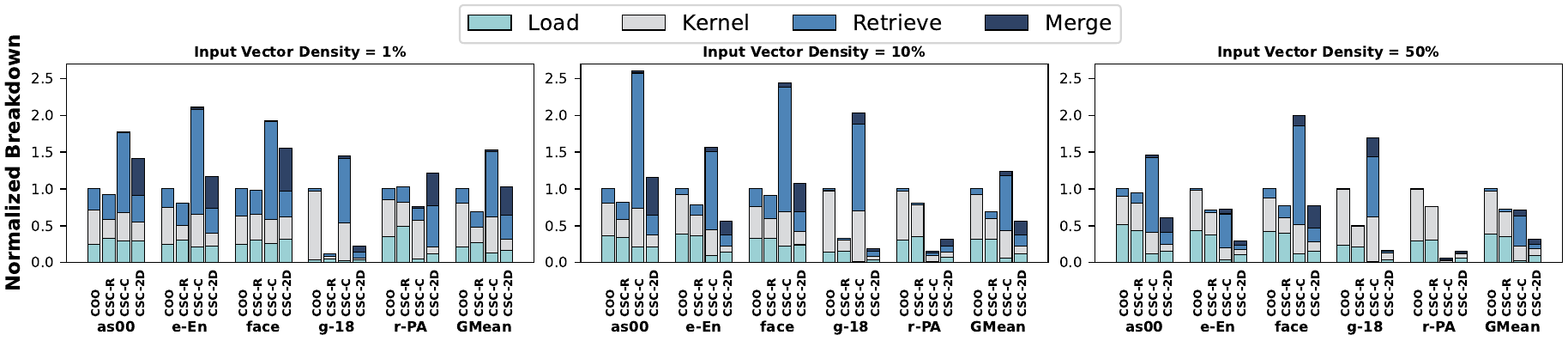}
  \caption{Execution time breakdown for SpMSpV variations using 2048 DPUs at input vector densities of $1\%$, $10\%$, and $50\%$ normalized to COO.}
\label{fig:spmspv-breakdown}
\end{figure*}

To understand the advantages and disadvantages of SpMSpV implementations based on input vector density, we compare various versions across several real-world datasets from different domains.
Figure~\ref{fig:spmspv-breakdown} compares four SpMSpV implementations: COO, CSC-R, CSC-C, and CSC-2D. The execution times are broken down by Load, Kernel, Retrieve, and Merge phases for input vector densities of $1\%$, $10\%$, and $50\%$. The x-axis represents the selected datasets and different implementation types, while the y-axis shows the normalized execution times relative to the COO version for each dataset.
For clarity, we display only a few representative datasets that show similar trends, as well as the normalized geometric mean of all implementations across all datasets. We omit density results between $1\%$ and $10\%$ and results between $10\%$ and $50\%$ as we observed very similar patterns. Additionally, we observed that CSR consistently performs worse than other SpMSpVs across various datasets and input vector densities. So, we exclude CSR as its performance was, on average, 2.8$\times$, 12.68$\times$, and 25.23$\times$ slower than the other versions at input vector densities of $1\%$, $10\%$, and $50\%$, respectively. SpMSpV performance varies depending on dataset features and input vector density. In particular, we make
four key observations.

\vspace{4pt}
$\blacktriangleright$ First, \textit{\textbf{CSC-2D} outperforms other SpMSpV versions across various datasets at higher input vector densities} due to lower DPU kernel times and partitioning both input and output vectors. The former will be further analyzed in Section~\ref{sec:kernel-analysis}. The latter reduces input vector load times compared to CSC-R and accelerates output vector retrieval compared to CSC-C.  For example, in the 'face' dataset at 50\% density, CSC-2D reduces input load by $7.3\times$ vs. CSC-R and output storage by $17.5\times$ vs. CSC-C.

\vspace{4pt}
$\blacktriangleright$ Second, \textit{\textbf{CSC-C} outperforms other versions for datasets like 'r-PA' across input vector densities}, especially due to lower retrieve times. The overall execution time in CSC-C is mainly determined by the time taken to retrieve the output vector, which is larger than in other CSC-based versions where each DPU transfers only a small portion of the output to the CPU. The 'r-PA' has small, uniform node degrees, so each DPU produces a small amount of output, which is efficiently compressed within the DPU, resulting in lower overall execution time for CSC-C compared to other formats.

\vspace{4pt}
$\blacktriangleright$ Third, \textit{at densities <10\%, CSC-2D is not always optimal.} For instance, \textbf{CSC-R} outperforms other versions for the 'g-18' dataset when the input vector density is below $10\%$. COO and CSC-R, which avoid output merging, can be more efficient than CSC-2D in such cases.

\vspace{4pt}
$\blacktriangleright$ Fourth, \textit{\textbf{COO} generally underperforms} due to processing the full adjacency matrix and transferring the entire input vector, but remains competitive in select cases.
\looseness=-1

\vspace{6pt}
\noindent
\fbox{
 \begin{minipage}{0.95\columnwidth}
\textbf{Summary:}
CSC format outperforms COO and CSR, but the optimal partitioning strategy depends on the input vector density and dataset characteristics.
 \end{minipage}%
}
\vspace{6pt}






\subsection{Effectiveness of Kernel Design}

\subsubsection{\textbf{SpMSpV vs. SpMV}}\label{sec_kernel}
\begin{figure}[!ht]
  \centering
  \includegraphics[width=1\linewidth]{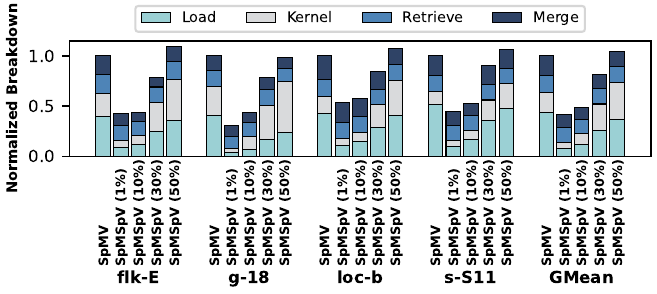}
  \caption{Comparison of execution time breakdown between the best-performing SpMV
  and SpMSpV at input vector densities of $1\%$, $10\%$, $30\%$, and $50\%$ using 2048 DPUs normalized to SpMV.}
\label{fig:spmspv-spmv}
\end{figure}

We evaluate how effectively we reduce the time associated with input vector loading using SpMSpV. Figure~\ref{fig:spmspv-spmv}
compares the best-performing SpMV from SparseP 
with the most efficient SpMSpV version (CSC-2D)
at input vector densities of $1\%$, $10\%$, $30\%$, and $50\%$. The execution times are categorized into Load, Kernel, Retrieve, and Merge phases. 
The y-axis displays normalized execution times relative to SpMV for each dataset. 
We report the normalized geometric mean of different implementations across all datasets in the last grouped bar. We make two main observations: \textcircled{1} \textit{SpMSpV significantly lowers input vector load time across all densities compared to SpMV}, with the largest gains below 30\%. Even at 50\%, it remains more efficient, and \textcircled{2} \textit{SpMSpV consistently outperforms SpMV version across most densities, and matches SpMV's performance at 50\% density}, showing robust effectiveness.

\vspace{4pt}
\subsubsection{\textbf{End-to-End Comparison: \graph~vs. SparseP}}
We compare the end-to-end performance of \graph~with the best-performing SpMV from SparseP, covering the full execution time per iteration, including data transfers, kernel computation, and result merging. While our earlier kernel-level comparison (Section \ref{sec_kernel}) focuses on single-iteration execution, this evaluation captures the complete multi-iteration behavior of graph applications.

\begin{figure*}[!ht]
  \centering
  \includegraphics[width=1\linewidth]{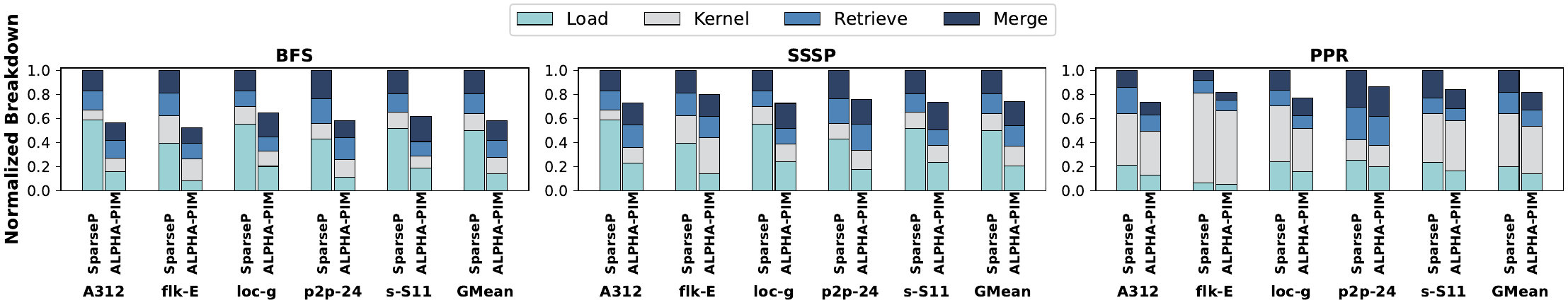}
  \caption{Performance Comparison: ALPHA-PIM vs. SparseP SpMV}
\label{fig:sparsep-alpha}
\end{figure*}

Figure~\ref{fig:sparsep-alpha} reports the normalized execution time of \graph~(with adaptive kernel switching) against SparseP’s SpMV baseline across three algorithms. \graph~achieves an average speedup of 1.72$\times$ for BFS, 1.34$\times$ for SSSP, and 1.22$\times$ for PPR, highlighting the practical benefit of adaptive switching over using SpMV exclusively.


\subsection{System-Level Comparison}
\subsubsection{\textbf{Impact of DPU Scaling on \graph~Performance}}\label{sec:dpu-scaling}
\begin{figure*}[!ht]
  \centering
  \includegraphics[width=1\linewidth]{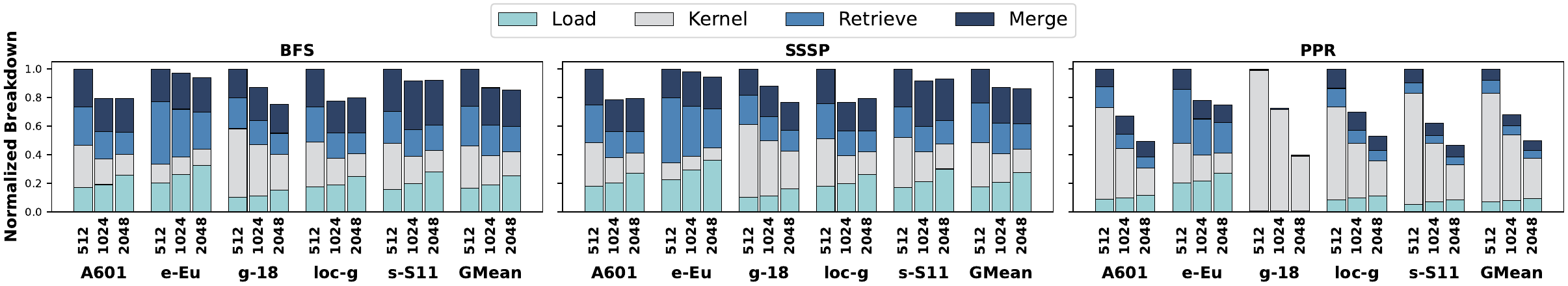}
  \caption{Execution time breakdown for BFS, SSSP, and PPR algorithms across varying DPU counts (512, 1024, 2048), normalized to 512 DPUs.}
\label{fig:evaluation-upmem}
\end{figure*}

Figure~\ref{fig:evaluation-upmem} shows the execution time breakdown of three graph algorithms—BFS, SSSP, and PPR—when varying the number of DPUs (512, 1024, 2048), normalized to 512. The execution times are broken down by Load, Kernel,
Retrieve, and Merge phases. We note that for these algorithms, each iteration includes a convergence check to determine if the algorithm should terminate. Therefore, we include the time spent on convergence checks in the merge time. 
The y-axis displays normalized execution times relative to 512 DPUs for each dataset. 
We also report the normalized geometric mean of different implementations across all datasets in the last grouped bar. We make three main observations.  

\vspace{4pt}
$\blacktriangleright$ First, \textit{in BFS and SSSP, much of the execution time goes to loading the input vector and retrieving the output}, due to their iterative nature. Each iteration involves a matrix-vector multiplication, where the input vector is the output vector from the previous iteration. Without inter-DPU communication, each DPU computes only a portion of the output, which must be sent back to the CPU, updated, and reloaded into the DPUs' MRAM for the next iteration.

\vspace{4pt}
$\blacktriangleright$ Second, in contrast, \textit{PPR’s execution is kernel-dominated.} PPR requires numerous floating-point operations, particularly multiplications, which are slow on UPMEM DPUs as they rely on software-emulated multiplication instead of faster hardware-based implementations.

\vspace{4pt}
$\blacktriangleright$ Third, \textit{using 2048 DPUs increases input vector load time, limiting speedup over 1024 DPUs} due to higher data transfer overhead. However, PPR benefits from more DPUs as its heavy kernel workload parallelizes well.

\vspace{6pt}
\noindent
\fbox{
 \begin{minipage}{0.95\columnwidth}
\textbf{Recommendations:}
The lack of inter-DPU communication leads to substantial vector transfer overhead between iterations, which could be mitigated by enabling direct interconnections. In algorithms like PPR, where kernel time dominates due to extensive floating-point operations, improved hardware/software support for floating-point is essential for better performance.
\end{minipage}%
}
\vspace{6pt}

\subsubsection{\textbf{UPMEM vs. CPU and GPU}}
We evaluate the UPMEM PIM system (Section \ref{sec:tools}) against an Intel CPU \cite{tools-cpu-gridgraph} and an NVIDIA RTX 3050 GPU \cite{tools-gpu}, focusing on latency, energy consumption, and compute utilization—defined as the achieved operations per second as a percentage of peak theoretical throughput. We report the compute utilization metric to evaluate the portion of the machine’s peak performance achieved on these applications in all systems.

This evaluation aims to assess the potential of UPMEM PIM as a general-purpose accelerator. Detailed micro-architectural specifications of the CPU and GPU are provided in Table \ref{table:all-real-machine-evaluation}. Peak performance for the CPU and GPU systems was calculated using peakperf \cite{tools-peakperf}, while the peak performance of the UPMEM PIM was determined using the method proposed in \cite{pim-sparsep}. The peak performance for the GPU, CPU, and UPMEM systems are 9.1 TFLOPS, 647.25 GFLOPS, and 4.66 GFLOPS, respectively.


For both CPU and GPU, we report results related to kernel execution without including data transfers between the host and GPU. In the case of UPMEM-based PIM, results are presented in two ways: first, reflecting kernel execution alone, and second, the total execution, which includes both DPU execution and inter-DPU synchronization. Energy consumption is measured using Intel RAPL \cite{tools-cpu-power} for the CPU, and NVIDIA SMI \cite{tools-gpu-power} for the GPU. For the UPMEM PIM system, energy is obtained from the DIMMs connected to the memory controllers via x86 sockets \cite{tools-pim-power}.

\begin{table}[hbt!]
\caption{CPU and GPU Micro-architectural Specs.}
\small
\label{table:all-real-machine-evaluation}
\resizebox{\linewidth}{!}{%
\begin{tabular}{|c|c|c|c|c|}
\hline
System & Total Cores   & Frequency & Capacity & Bandwidth  \\ \hline
Intel i7-1265U& 10 (12 threads) & 1.8 GHz &  64GB & 83.2 GB/s \\ \hline
NVIDIA RTX 3050 & 2560 CUDA cores & 1.55 GHz & 8GB  & 224 GB/s 
\\ \hline
\end{tabular}
}
\end{table}

We utilize the GridGraph library \cite{compare-cpu-gridgraph} for CPU-based processing and employ cuGraph \cite{app-cugraph} from the RAPIDS suite \cite{app-RAPIDS} on the GPU, which incorporates algorithms from NVIDIA’s widely-used nvGRAPH library \cite{app-nvgraph} for efficient graph analytics. 
Table \ref{tab:cpu-gpu-comparision} presents the execution time (in milliseconds), energy consumption (in joules), and compute utilization (as a percentage) for three algorithms across all iterations on selected datasets for each system. We derive three key observations.

\vspace{4pt}
$\blacktriangleright$ First, \graph~\textit{achieves an average speedup of 10.2$\times$ (kernel execution time) and 2.6$\times$ (total time) for BFS, 48.8$\times$ and 10.4$\times$ for SSSP, and 3.6$\times$ and 1.7$\times$ for PPR over a conventional CPU baseline.}


\vspace{4pt}
$\blacktriangleright$ Second, \graph~\textit{shows much higher compute utilization with UPMEM averaging: 18.5\% (kernel) / 5\% (total) for BFS, 31.8\% / 7.1\% for SSSP, and 6.4\% / 2.8\% for PPR}, compared to only 0.01–0.05\% on CPU/GPU. This suggests that the PIM system is better balanced between memory and compute resources for SpMV/SpMSpV operations. While CPUs and GPUs often face limitations due to memory bandwidth, PIM systems are more constrained by their computing capabilities, highlighting an opportunity for further improvements. 

\vspace{4pt}
$\blacktriangleright$ Third, \textit{the GPU outperforms  CPU and UPMEM in execution time and energy efficiency}, owing to its mature, highly optimized parallel architecture. UPMEM, by contrast, faces two major architectural challenges: \textcircled{1} the lack of direct inter-DPU communication, which requires repeated CPU-mediated data transfers—especially costly in iterative graph workloads where input and output vectors are exchanged in every iteration (see Section~\ref{sec:dpu-scaling}); and \textcircled{2} a distributed DRAM architecture without global memory sharing, which prevents data reuse across DPUs and increases memory movement overhead. Additionally, as we will analyze in Section~\ref{sec:kernel-analysis}, UPMEM suffers from underutilized DPU kernel execution, with low IPC and idle cycles caused by structural hazards and synchronization overheads. While PIM shows great promise for near-data processing, these bottlenecks currently limit its competitiveness. Our goal is to deeply characterize these issues and inform future PIM hardware and software optimizations.

\begin{table*}[t]
\centering
\caption{Execution Time (ms), Compute Utilization (\%), and Energy Consumption (J) in CPU, GPU, and PIM for different datasets}
\label{tab:cpu-gpu-comparision}
\resizebox{\textwidth}{!}{
\begin{tabular}{|c|c|cccccc|cccccc|cccccc|}
\hline
\multirow{2}{*}{\textbf{Algo.}} & \multirow{2}{*}{\textbf{System}} & \multicolumn{6}{c|}{\textbf{Execution Time (ms)}} & \multicolumn{6}{c|}{\textbf{Compute Utilization (\%)}} & \multicolumn{6}{c|}{\textbf{Energy Consumption (J)}} \\ \cline{3-20}
 & & \textbf{A302} & \textbf{as00} & \textbf{s-S11} & \textbf{p2p-24} & \textbf{e-En} & \textbf{face} & \textbf{A302} & \textbf{as00} & \textbf{s-S11} & \textbf{p2p-24} & \textbf{e-En} & \textbf{face} & \textbf{A302} & \textbf{as00} & \textbf{s-S11} & \textbf{p2p-24} & \textbf{e-En} & \textbf{face} \\
\hline
\multirow{4}{*}{\textbf{BFS}} 
 & \textbf{CPU}          & 541.1 & 38.5 & 44.5 & 117.1 & 44.5 & 27.1 & 0.09 & 0.002 & 0.07 & 0.001 & 0.04 & 0.02 & 17.30 & 0.90 & 1.05 & 3.18 & 1.10 & 0.64 \\
 & \textbf{GPU}          & 7.08  & 0.89 & 2.2  & 1.23  & 1.22 & 0.96 & 0.50 & 0.01  & 0.11 & 0.01  & 0.09 & 0.04 & 0.14  & 0.02 & 0.05 & 0.03 & 0.02 & 0.02 \\
 & \textbf{UPMEM-Kernel} & 76.6  & 2.62 & 8.2  & 5.67  & 8.24 & 3.53 & 90.1 & 3.58  & 55.7 & 3.81  & 26.8 & 22.2 & 35.6  & 1.22 & 3.80 & 2.63 & 3.82 & 1.64 \\
 & \textbf{UPMEM-Total}  & 241.1 & 13.3 & 33.4 & 23.0  & 31.5 & 9.55 & 28.6 & 0.71  & 13.7 & 0.94  & 7.01 & 8.20 & 111.9 & 6.15 & 15.5 & 10.7 & 14.6 & 4.43 \\
\hline
\multirow{4}{*}{\textbf{SSSP}} 
 & \textbf{CPU}          & 1900  & 61   & 1056 & 166.5 & 656.1 & 232  & 0.04 & 0.002 & 0.01 & 0.002 & 0.01 & 0.01 & 57.2  & 1.52 & 24.5 & 4.74 & 13.6 & 6.44 \\
 & \textbf{GPU}          & 12.7  & 13.0 & 12.9 & 12.8  & 12.5  & 13.1 & 0.42 & 0.0007& 0.04 & 0.002 & 0.01 & 0.02 & 0.25  & 0.26 & 0.26 & 0.26 & 0.25 & 0.28 \\
 & \textbf{UPMEM-Kernel} & 62.7  & 4.3  & 8.3  & 7.9   & 11.8  & 5.3  & 165  & 4.36  & 131  & 6.98  & 39.5 & 40.1 & 29.1  & 0.20 & 3.84 & 3.67 & 5.46 & 2.46 \\
 & \textbf{UPMEM-Total}  & 340   & 19.9 & 49.3 & 29.9  & 43.3  & 20.2 & 30.5 & 0.94  & 21.9 & 1.85  & 10.7 & 10.5 & 158   & 9.23 & 22.9 & 13.9 & 20.1 & 9.36 \\
\hline
\multirow{4}{*}{\textbf{PPR}} 
 & \textbf{CPU}          & 216   & 126  & 177  & 88.5  & 197   & 84.0 & 0.13 & 0.002 & 0.02 & 0.004 & 0.02 & 0.01 & 7.25  & 3.46 & 4.48 & 2.22 & 5.63 & 2.40 \\
 & \textbf{GPU}          & 18.2  & 14.3 & 18.6 & 13.0  & 18.0  & 12.7 & 0.11 & 0.002 & 0.02 & 0.003 & 0.01 & 0.01 & 0.35  & 0.28 & 0.33 & 0.27 & 0.31 & 0.26 \\
 & \textbf{UPMEM-Kernel} & 78.5  & 37.2 & 76.5 & 17.7  & 58.7  & 22.4 & 51.4 & 0.90  & 10.6 & 2.71  & 8.66 & 6.49 & 36.6  & 13.8 & 17.3 & 26.3 & 27.4 & 10.1 \\
 & \textbf{UPMEM-Total}  & 196.2 & 45.9 & 144  & 46.9  & 84.4  & 104  & 20.5 & 0.73  & 3.97 & 1.22  & 6.09 & 1.40 & 91.0  & 21.3 & 66.7 & 21.8 & 38.7 & 48.6 \\
\hline
\end{tabular}}
\end{table*}

\subsection{In-Depth Kernel Bottleneck Analysis in \graph}\label{sec:kernel-analysis}
We previously observed that GPUs outperform UPMEM due to their superior computational throughput, while UPMEM, as a newer technology, requires further optimization to achieve competitive performance. This section examines architectural bottlenecks in SpMSpV and SpMV kernels that hinder the performance of graph applications on UPMEM. We analyzed detailed profiling metrics obtained from PIMulator \cite{tools-upimulator}. Using PIMulator, we identify key microarchitectural bottlenecks in UPMEM-PIM that contribute to performance loss, providing insights into kernel behavior that are difficult to obtain from real-system experiments.
\vspace{4pt}
\subsubsection{\textbf{DPU Utilization and Thread-Level Activity}}
\begin{figure}[!ht]
  \centering
  \includegraphics[width=1\linewidth]{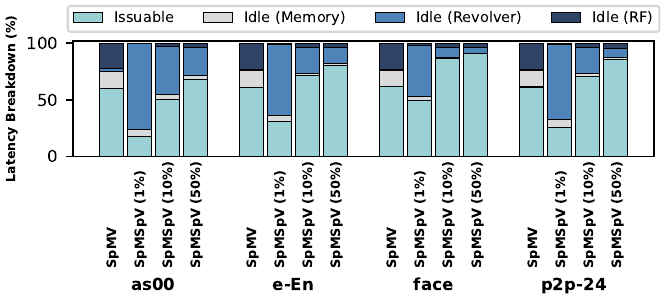}
  \caption{DPU runtime breakdown into active (green) and idle (non-green bars) cycles for SpMV and SpMSpV at input vector densities of  $1\%$, $10\%$, and $50\%$. Idle cycles are categorized by stall reasons: memory (gray), revolver pipeline (light blue), and RF structural hazard (dark blue).}
\label{fig:spmspv-stalls}
\end{figure}

Figure~\ref{fig:spmspv-stalls} breaks down DPU execution into: \textcircled{1} periods when the thread scheduler actively issues threads into the pipeline (green bar); and \textcircled{2} periods when the scheduler is idle (non-green bars), caused by threads waiting on memory operations, pipeline scheduling constraints, or structural hazards in the register file. These factors limit DPU utilization, leading to low IPC. We make four key observations.

\vspace{4pt}
$\blacktriangleright$ First, \textit{SpMSpV with input densities $> 10\%$ shows a higher percentage of issuable cycles than SpMV.}
This is because SpMSpV focuses only on the non-zero entries of the input vector, reducing unnecessary computation and improving memory access locality. This results in fewer stalls and better DPU pipeline utilization, allowing more instructions to be issued per cycle.

\vspace{4pt}
$\blacktriangleright$ Second,\textit{ as input density increases in SpMSpV (from 1\% to 50\%), revolver pipeline stalls decrease.}
Higher input density means more non-zero elements are available per active column, enabling better instruction-level parallelism. This reduces scheduling gaps in UPMEM’s revolver pipeline, which requires 11 cycles between dispatching consecutive instructions from the same thread.

\vspace{4pt}
$\blacktriangleright$ Third, \textit{SpMV suffers from more memory and register file (RF) stalls compared to SpMSpV.}
SpMV accesses the entire input vector—regardless of actual sparsity—leading to irregular and less predictable memory access patterns, which increase memory stalls. Additionally, since every matrix element contributes to output updates, SpMV generates more RF contention, whereas SpMSpV avoids many such operations by skipping zero elements.

\vspace{4pt}
$\blacktriangleright$ Fourth,\textit{ at 1\% density, SpMSpV exhibits higher revolver pipeline stalls compared to higher densities.} At low density, fewer active columns lead to reduced useful computation per thread and more frequent synchronization on shared output vector entries. In CSC-based SpMSpV, mutexes serialize these updates, preventing multiple tasklets from progressing simultaneously, which underutilizes the pipeline despite having sufficient tasklets. This combination of low computational load and contention leads to elevated revolver stalls.

\begin{figure}[!ht]
  \centering
  \includegraphics[width=1\linewidth]{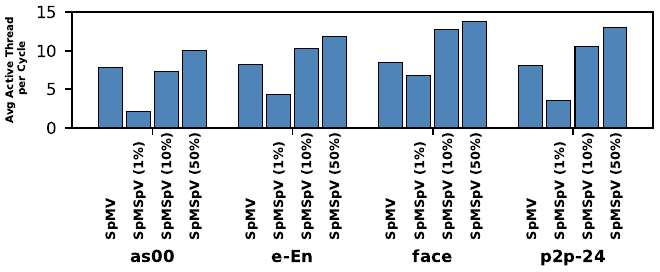}
  \caption{Average Active Thread Per Cycle}
\label{fig:active-thread}
\end{figure}

Figure~\ref{fig:active-thread} presents the average number of active threads per cycle for SpMV and SpMSpV kernels across various datasets and input vector densities. \emph{SpMSpV achieves higher thread activity as input density increases, benefiting from more parallel work per DPU.} At 1\% density, limited active columns reduce thread engagement, but this improves at 10\% and 50\%. In contrast, SpMV shows lower thread activity due to irregular input access, reinforcing its lower DPU utilization observed earlier.



\vspace{6pt}
\noindent
\fbox{%
  \begin{minipage}{0.95\columnwidth}
\textbf{Recommendations:}
Revolver pipeline efficiency could be improved by adding intra-thread data forwarding for independent instructions, as proposed in \cite{tools-upimulator}. 
Further, enabling non-blocking DMA would let tasklets continue computation while waiting for memory, requiring changes to the pipeline’s thread dispatch logic to improve instruction-level parallelism.
 \end{minipage}%
}
\vspace{6pt}

\subsubsection{\textbf{Instruction Mix and Synchronization Overheads}}
\begin{figure}[!ht]
  \centering
  \includegraphics[width=1\linewidth]{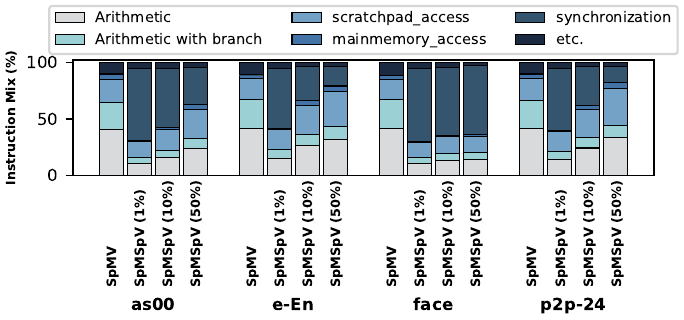}
  \caption{Instruction mix for SpMV and SpMSpV at input vector densities of $1\%$, $10\%$, and $50\%$.}
\label{fig:spmspv-instmix}
\end{figure}

Figure~\ref{fig:spmspv-instmix} presents the instruction mix across SpMV and SpMSpV at $1\%$, $10\%$, and $50\%$ input vector densities. Three key insights follow:

\vspace{4pt}
$\blacktriangleright$ First, \textit{synchronization instructions (e.g., barrier, mutex\_lock) account for a large share in SpMSpV, especially at low densities.} Sparse input leads to fewer active updates and higher contention over shared output entries, increasing synchronization. Higher densities distribute updates more evenly, reducing contention and synchronization overhead.

\vspace{4pt}
$\blacktriangleright$ Second, S\textit{pMV exhibits more arithmetic operations than SpMSpV}, as it processes all rows regardless of input vector sparsity, while SpMSpV skips inactive columns.


\vspace{4pt}
$\blacktriangleright$ Third, s\textit{cratchpad instructions account for a non-trivial share across all implementations due to UPMEM’s scratchpad-centric model}, which requires explicit DRAM-to-WRAM transfers as all register operands must be loaded from WRAM. To mitigate this overhead, UPMEM executes Load/Store instructions with single-cycle latency \cite{tools-upimulator}, preserving efficiency despite the lack of automated caching.

\vspace{6pt}
\noindent
\fbox{
 \begin{minipage}{0.95\columnwidth}
\textbf{Recommendations:}
Implementing more efficient synchronization techniques could reduce idle time and improve system efficiency.
 \end{minipage}%
}
\vspace{6pt}




\section{Related Work} \label{sec:related-work}

To our knowledge, this is the first study to extensively characterize linear-algebraic graph processing on a real PIM system. 
We provide a brief overview of related prior work.

\vspace{4pt}
\noindent\textbf{Graph Processing in PIM Systems.}
PIM-based accelerators boost graph processing by improving memory bandwidth efficiency, integrating computation directly within memory to reduce data transfers and mitigate bandwidth bottlenecks
\cite{accelerator-pim-scalable,accelerator-pim-graphp,accelerator-pim-graphh,accelerator-pim-graphpim, accelerator-pim-sisa, accelerator-pim-pim-enabled, accelerator-pim-conda, accelerator-pim-lazy, accelerator-pim-graphq, related-pimpam, related-gaas, related-psyncpim}.

\vspace{4pt}
\noindent\textbf{Sparse Computation in PIM Systems.} Giannoula et al. \cite{pim-sparsep} present the first SpMV library for real PIM systems. Sun et al. \cite{sparse-pim-Abc} design buffer-optimized DIMMs enabling inter-DIMM communication for irregular tasks like SpMV and graph processing. Fujiki et al. \cite{sparse-pim-gpu} enhance GPU memory controllers by integrating PIM cores to convert matrices into DCSR format. Xie et al. \cite{sparse-pim-spaceA} propose a 3D PNM design for SpMV using heterogeneous PIM in HMC. Zhu et al. \cite{sparse-pim-3d-stacked} develop a PIM accelerator for SpMM in 3D-stacked memory systems.

\vspace{4pt}
\noindent\textbf{Graph Processing in Commodity Systems.}
Several CPU-based frameworks support either sequential or coarse-grained parallel programming models. To address scalability challenges, distributed CPU-based frameworks have been developed to tackle issues such as synchronization overhead and load imbalance \cite{frameworks-cpu-bridging,frameworks-cpu-graphchi,frameworks-cpu-graphx,frameworks-cpu-Green-Marl,frameworks-cpu-help,frameworks-cpu-lightweight,frameworks-cpu-powergraph,frameworks-cpu-pregel}. Several GPU-based frameworks implement graph primitives, with specialized parallel graph algorithms \cite{frameworks-single-accelerating,framewoks-single-scalablebc,frameworks-single-betweenness,frameworks-single-bfs-4k,frameworks-single-comparison,frameworks-single-computingBC,frameworks-single-connectivity,frameworks-single-delta,frameworks-single-direction,frameworks-single-enterprise,frameworks-single-fasterBC,frameworks-single-fasttc,frameworks-single-maximumwarp,frameworks-single-scalable,frameworks-single-worksssp,frameworks-single-computingcc,frameworks-single-ibfs,frameworks-single-counting,frameworks-single-fastpr, frameworks-single-almasri2023parallelizing, frameworks-single-almasri2022parallel, frameworks-single-yamout2022parallel, frameworks-single-diab2020ktrussexplorer, frameworks-single-olabi2022compiler}. However, these frameworks do not generalize well across different types of graph applications but show significant performance gains. Generalized frameworks \cite{frameworks-highlevel-cusha,frameworks-highlevel-efficient,frameworks-highlevel-garaph,frameworks-highlevel-graphie,frameworks-highlevel-gunrock,frameworks-highlevel-mapgraph,frameworks-highlevel-medusa,frameworks-highlevel-optimization,frameworks-highlevel-sep,frameworks-highlevel-simd,frameworks-highlevel-subway,frameworks-highlevel-tigr,frameworks-highlevel-BGL,frameworks-highlevel-compiler,frameworks-highlevel-gpudistributed,frameworks-highlevel-gts,frameworks-highlevel-groute,frameworks-highlevel-digraph,frameworks-highlevel-emptyheaded,frameworks-highlevel-multigraph,frameworks-highlevel-yoke,frameworks-highlevel-gstream,frameworks-highlevel-ligra,frameworks-highlevel-medusa,frameworks-highlevel-efficientcpu-gpu,frameworks-highlevel-gbtl,frameworks-highlevel-towards,frameworks-highlevel-scalablesimd} offer greater programming flexibility. Studies on GPU-optimized SpMV \cite{sparse-framework-biell,sparse-framework-fastsparse,sparse-framework-implementingsparse,sparse-framework-performance,sparse-framework-sparsematrix,sparse-framework-yaSpMV,accelerator-algebra-efficient,motivation-spmspv-fast,accelerator-algebra-slimsell} focus on efficient matrix formats (e.g., CSR, COO). Tools like ClSpMV \cite{sparse-clspmv} and ML methods \cite{background-csr} help select formats, while libraries such as MKL \cite{sparse-lib-mkl} and cuSPARSE \cite{cusparse} are widely used, with some research exploring tensor cores \cite{sparse-tensorcore}.

\section{Conclusion}\label{sec:conclusion}

We introduce \graph, the first 
linear-algebraic graph application framework designed for a real PIM system, and conduct a comprehensive analysis of widely-used graph algorithms on UPMEM. 
Our study underscores the crucial role of selecting effective partitioning strategies and compressed matrix formats to enhance performance on PIM systems.
We demonstrate the critical need for PIM architectures incorporating direct inter-DPU communication capabilities to address data transfer challenges in graph applications, paving the way for future advancements in PIM technology for large-scale graph processing tasks.


\section*{Acknowledgments}
We acknowledge the generous support of our industrial partners, including Intel, Ampere Computing, Google, Huawei, and Microsoft. This work was supported in part by the Natural Sciences and Engineering Research Council of Canada (NSERC); an SFU Faculty Recruitment Grant; the Semiconductor Research Corporation (SRC); the ETH Future Computing Laboratory (EFCL); the AI Chip Center for Emerging Smart Systems (ACCESS), sponsored by InnoHK funding, Hong Kong SAR; and the European Union’s Horizon research and innovation programme. We are grateful to UPMEM for granting us access to their systems. We also thank the anonymous reviewers for their constructive feedback.

\balance
\bibliographystyle{IEEEtran}
\bibliography{ref}

@inproceedings{frameworks-cpu-pregel,
  title={Pregel: a system for large-scale graph processing},
  author={Malewicz, Grzegorz and Austern, Matthew H and Bik, Aart JC and Dehnert, James C and Horn, Ilan and Leiser, Naty and Czajkowski, Grzegorz},
  booktitle={Proceedings of the 2010 ACM SIGMOD International Conference on Management of data},
  pages={135--146},
  year={2010}
}

@article{frameworks-cpu-bridging,
  title={A bridging model for parallel computation},
  author={Valiant, Leslie G},
  journal={Communications of the ACM},
  volume={33},
  number={8},
  pages={103--111},
  year={1990},
  publisher={ACM New York, NY, USA}
}

@inproceedings{frameworks-cpu-powergraph,
  title={Powergraph: Distributed graph-parallel computation on natural graphs},
  author={Gonzalez, Joseph E and Low, Yucheng and Gu, Haijie and Bickson, Danny and Guestrin, Carlos},
  booktitle={Presented as part of the 10th $\{$USENIX$\}$ Symposium on Operating Systems Design and Implementation ($\{$OSDI$\}$ 12)},
  pages={17--30},
  year={2012}
}

@inproceedings{frameworks-cpu-graphchi,
  title={Graphchi: Large-scale graph computation on just a $\{$PC$\}$},
  author={Kyrola, Aapo and Blelloch, Guy and Guestrin, Carlos},
  booktitle={Presented as part of the 10th $\{$USENIX$\}$ Symposium on Operating Systems Design and Implementation ($\{$OSDI$\}$ 12)},
  pages={31--46},
  year={2012}
}

@inproceedings{frameworks-cpu-lightweight,
  title={A lightweight infrastructure for graph analytics},
  author={Nguyen, Donald and Lenharth, Andrew and Pingali, Keshav},
  booktitle={Proceedings of the Twenty-Fourth ACM Symposium on Operating Systems Principles},
  pages={456--471},
  year={2013}
}

@inproceedings{frameworks-cpu-Green-Marl,
  title={Green-Marl: a DSL for easy and efficient graph analysis},
  author={Hong, Sungpack and Chafi, Hassan and Sedlar, Edic and Olukotun, Kunle},
  booktitle={Proceedings of the seventeenth international conference on Architectural Support for Programming Languages and Operating Systems},
  pages={349--362},
  year={2012}
}

@inproceedings{frameworks-cpu-graphx,
  title={Graphx: Graph processing in a distributed dataflow framework},
  author={Gonzalez, Joseph E and Xin, Reynold S and Dave, Ankur and Crankshaw, Daniel and Franklin, Michael J and Stoica, Ion},
  booktitle={11th $\{$USENIX$\}$ Symposium on Operating Systems Design and Implementation ($\{$OSDI$\}$ 14)},
  pages={599--613},
  year={2014}
}

@inproceedings{frameworks-cpu-help,
  title={HelP: High-level primitives for large-scale graph processing},
  author={Salihoglu, Semih and Widom, Jennifer},
  booktitle={Proceedings of Workshop on GRAph Data management Experiences and Systems},
  pages={1--6},
  year={2014}
}

@inproceedings{frameworks-single-accelerating,
  title={Accelerating large graph algorithms on the GPU using CUDA},
  author={Harish, Pawan and Narayanan, Petter J},
  booktitle={International conference on high-performance computing},
  pages={197--208},
  year={2007},
  organization={Springer}
}

@article{frameworks-single-maximumwarp,
  title={Accelerating CUDA graph algorithms at maximum warp},
  author={Hong, Sungpack and Kim, Sang Kyun and Oguntebi, Tayo and Olukotun, Kunle},
  journal={Acm Sigplan Notices},
  volume={46},
  number={8},
  pages={267--276},
  year={2011},
  publisher={ACM New York, NY, USA}
}

@article{frameworks-single-scalable,
  title={Scalable GPU graph traversal},
  author={Merrill, Duane and Garland, Michael and Grimshaw, Andrew},
  journal={ACM Sigplan Notices},
  volume={47},
  number={8},
  pages={117--128},
  year={2012},
  publisher={ACM New York, NY, USA}
}

@inproceedings{frameworks-single-direction,
  title={Ligra breadth-first search},
  author={Beamer, Scott and Asanovic, Krste and Patterson, David},
  booktitle={SC'12: Proceedings of the International Conference on High Performance Computing, Networking, Storage and Analysis},
  pages={1--10},
  year={2012},
  organization={IEEE}
}

@inproceedings{frameworks-single-enterprise,
  title={Enterprise: breadth-first graph traversal on GPUs},
  author={Liu, Hang and Huang, H Howie},
  booktitle={Proceedings of the International Conference for High Performance Computing, Networking, Storage and Analysis},
  pages={1--12},
  year={2015}
}

@article{frameworks-single-bfs-4k,
  title={BFS-4K: an efficient implementation of BFS for kepler GPU architectures},
  author={Busato, Federico and Bombieri, Nicola},
  journal={IEEE Transactions on Parallel and Distributed Systems},
  volume={26},
  number={7},
  pages={1826--1838},
  year={2014},
  publisher={IEEE}
}

@inproceedings{frameworks-single-connectivity,
  title={A fast GPU algorithm for graph connectivity},
  author={Soman, Jyothish and Kishore, Kothapalli and Narayanan, PJ},
  booktitle={2010 IEEE International Symposium on Parallel \& Distributed Processing, Workshops and Phd Forum (IPDPSW)},
  pages={1--8},
  year={2010},
  organization={IEEE}
}

@inproceedings{frameworks-single-comparison,
  title={A comparison of parallel algorithms for connected components},
  author={Greiner, John},
  booktitle={Proceedings of the sixth annual ACM symposium on Parallel algorithms and architectures},
  pages={16--25},
  year={1994}
}

@article{frameworks-single-fasterBC,
  title={A faster algorithm for betweenness centrality},
  author={Brandes, Ulrik},
  journal={Journal of mathematical sociology},
  volume={25},
  number={2},
  pages={163--177},
  year={2001},
  publisher={Taylor \& Francis}
}

@inproceedings{frameworks-single-computingBC,
  title={Computing betweenness centrality for small world networks on a GPU},
  author={Pande, P and Bader, David A},
  booktitle={15th Annual High performance embedded computing workshop (HPEC)},
  year={2011}
}

@inproceedings{frameworks-single-betweenness,
  title={Betweenness centrality on GPUs and heterogeneous architectures},
  author={Sariy{\"u}ce, Ahmet Erdem and Kaya, Kamer and Saule, Erik and {\c{C}}ataly{\"u}rek, {\"U}mit V},
  booktitle={Proceedings of the 6th Workshop on General Purpose Processor Using Graphics Processing Units},
  pages={76--85},
  year={2013}
}

@inproceedings{framewoks-single-scalablebc,
  title={Scalable and high performance betweenness centrality on the GPU},
  author={McLaughlin, Adam and Bader, David A},
  booktitle={SC'14: Proceedings of the International Conference for High Performance Computing, Networking, Storage and Analysis},
  pages={572--583},
  year={2014},
  organization={IEEE}
}

@inproceedings{frameworks-single-worksssp,
  title={Work-efficient parallel GPU methods for single-source shortest paths},
  author={Davidson, Andrew and Baxter, Sean and Garland, Michael and Owens, John D},
  booktitle={2014 IEEE 28th International Parallel and Distributed Processing Symposium},
  pages={349--359},
  year={2014},
  organization={IEEE}
}

@article{frameworks-single-delta,
  title={$\Delta$-stepping: a parallelizable shortest path algorithm},
  author={Meyer, Ulrich and Sanders, Peter},
  journal={Journal of Algorithms},
  volume={49},
  number={1},
  pages={114--152},
  year={2003},
  publisher={Elsevier}
}

@inproceedings{frameworks-single-fasttc,
  title={Fast triangle counting on the GPU},
  author={Green, Oded and Yalamanchili, Pavan and Mungu{\'\i}a, Llu{\'\i}s-Miquel},
  booktitle={Proceedings of the 4th Workshop on Irregular Applications: Architectures and Algorithms},
  pages={1--8},
  year={2014}
}

@inproceedings{frameworks-single-computingcc,
  title={Computing strongly connected components in parallel on CUDA},
  author={Barnat, Jiri and Bauch, Petr and Brim, Lubos and Ce{\v{s}}ka, Milan},
  booktitle={2011 IEEE International Parallel \& Distributed Processing Symposium},
  pages={544--555},
  year={2011},
  organization={IEEE}
}

@inproceedings{frameworks-single-ibfs,
  title={ibfs: Concurrent breadth-first search on gpus},
  author={Liu, Hang and Huang, H Howie and Hu, Yang},
  booktitle={Proceedings of the 2016 International Conference on Management of Data},
  pages={403--416},
  year={2016}
}

@inproceedings{frameworks-single-counting,
  title={Counting triangles in large graphs on GPU},
  author={Polak, Adam},
  booktitle={2016 IEEE International Parallel and Distributed Processing Symposium Workshops (IPDPSW)},
  pages={740--746},
  year={2016},
  organization={IEEE}
}

@inproceedings{frameworks-single-fastpr,
  title={Fast pagerank computation on a gpu cluster},
  author={Rungsawang, Arnon and Manaskasemsak, Bundit},
  booktitle={2012 20th Euromicro International Conference on Parallel, Distributed and Network-based Processing},
  pages={450--456},
  year={2012},
  organization={IEEE}
}

@inproceedings{frameworks-single-almasri2023parallelizing,
  title={Parallelizing maximal clique enumeration on gpus},
  author={Almasri, Mohammad and Chang, Yen-Hsiang and El Hajj, Izzat and Nagi, Rakesh and Xiong, Jinjun and Hwu, Wen-mei},
  booktitle={2023 32nd International Conference on Parallel Architectures and Compilation Techniques (PACT)},
  pages={162--175},
  year={2023},
  organization={IEEE}
}

@inproceedings{frameworks-single-almasri2022parallel,
  title={Parallel k-clique counting on gpus},
  author={Almasri, Mohammad and Hajj, Izzat El and Nagi, Rakesh and Xiong, Jinjun and Hwu, Wen-mei},
  booktitle={Proceedings of the 36th ACM International Conference on Supercomputing},
  pages={1--14},
  year={2022}
}

@inproceedings{frameworks-single-yamout2022parallel,
  title={Parallel vertex cover algorithms on gpus},
  author={Yamout, Peter and Barada, Karim and Jaljuli, Adnan and Mouawad, Amer E and El Hajj, Izzat},
  booktitle={2022 IEEE International Parallel and Distributed Processing Symposium (IPDPS)},
  pages={201--211},
  year={2022},
  organization={IEEE}
}

@inproceedings{frameworks-single-diab2020ktrussexplorer,
  title={Ktrussexplorer: Exploring the design space of k-truss decomposition optimizations on gpus},
  author={Diab, Safaa and Olabi, Mhd Ghaith and El Hajj, Izzat},
  booktitle={2020 IEEE High Performance Extreme Computing Conference (HPEC)},
  pages={1--8},
  year={2020},
  organization={IEEE}
}

@inproceedings{frameworks-single-olabi2022compiler,
  title={A compiler framework for optimizing dynamic parallelism on GPUs},
  author={Olabi, Mhd Ghaith and Luna, Juan G{\'o}mez and Mutlu, Onur and Hwu, Wen-mei and El Hajj, Izzat},
  booktitle={2022 IEEE/ACM International Symposium on Code Generation and Optimization (CGO)},
  pages={1--13},
  year={2022},
  organization={IEEE}
}

@inproceedings{frameworks-highlevel-graphie,
  title={Graphie: Large-scale asynchronous graph traversals on just a gpu},
  author={Han, Wei and Mawhirter, Daniel and Wu, Bo and Buland, Matthew},
  booktitle={2017 26th International Conference on Parallel Architectures and Compilation Techniques (PACT)},
  pages={233--245},
  year={2017},
  organization={IEEE}
}

@inproceedings{frameworks-highlevel-subway,
  title={Subway: minimizing data transfer during out-of-GPU-memory graph processing},
  author={Sabet, Amir Hossein Nodehi and Zhao, Zhijia and Gupta, Rajiv},
  booktitle={Proceedings of the Fifteenth European Conference on Computer Systems},
  pages={1--16},
  year={2020}
}

@inproceedings{frameworks-highlevel-simd,
  title={Simd-x: Programming and processing of graph algorithms on gpus},
  author={Liu, Hang and Huang, H Howie},
  booktitle={2019 $\{$USENIX$\}$ Annual Technical Conference ($\{$USENIX$\}$$\{$ATC$\}$ 19)},
  pages={411--428},
  year={2019}
}

@inproceedings{frameworks-highlevel-garaph,
  title={Garaph: Efficient GPU-accelerated graph processing on a single machine with balanced replication},
  author={Ma, Lingxiao and Yang, Zhi and Chen, Han and Xue, Jilong and Dai, Yafei},
  booktitle={2017 $\{$USENIX$\}$ Annual Technical Conference ($\{$USENIX$\}$$\{$ATC$\}$ 17)},
  pages={195--207},
  year={2017}
}

@article{frameworks-highlevel-tigr,
  title={Tigr: Transforming irregular graphs for gpu-friendly graph processing},
  author={Nodehi Sabet, Amir Hossein and Qiu, Junqiao and Zhao, Zhijia},
  journal={ACM SIGPLAN Notices},
  volume={53},
  number={2},
  pages={622--636},
  year={2018},
  publisher={ACM New York, NY, USA}
}

@inproceedings{frameworks-highlevel-sep,
  title={SEP-graph: finding shortest execution paths for graph processing under a hybrid framework on GPU},
  author={Wang, Hao and Geng, Liang and Lee, Rubao and Hou, Kaixi and Zhang, Yanfeng and Zhang, Xiaodong},
  booktitle={Proceedings of the 24th Symposium on Principles and Practice of Parallel Programming},
  pages={38--52},
  year={2019}
}

@article{frameworks-highlevel-medusa,
  title={Medusa: Simplified graph processing on GPUs},
  author={Zhong, Jianlong and He, Bingsheng},
  journal={IEEE Transactions on Parallel and Distributed Systems},
  volume={25},
  number={6},
  pages={1543--1552},
  year={2013},
  publisher={IEEE}
}

@article{frameworks-highlevel-efficient,
  title={Efficient large-scale graph processing on hybrid CPU and GPU systems},
  author={Gharaibeh, Abdullah and Reza, Tahsin and Santos-Neto, Elizeu and Costa, Lauro Beltrao and Sallinen, Scott and Ripeanu, Matei},
  journal={arXiv preprint arXiv:1312.3018},
  year={2013}
}

@article{frameworks-highlevel-optimization,
  title={Optimization of asynchronous graph processing on GPU with hybrid coloring model},
  author={Shi, Xuanhua and Liang, Junling and Di, Sheng and He, Bingsheng and Jin, Hai and Lu, Lu and Wang, Zhixiang and Luo, Xuan and Zhong, Jianlong},
  journal={ACM SIGPLAN Notices},
  volume={50},
  number={8},
  pages={271--272},
  year={2015},
  publisher={ACM New York, NY, USA}
}

@inproceedings{frameworks-highlevel-mapgraph,
  title={Mapgraph: A high level api for fast development of high performance graph analytics on gpus},
  author={Fu, Zhisong and Personick, Michael and Thompson, Bryan},
  booktitle={Proceedings of Workshop on GRAph Data management Experiences and Systems},
  pages={1--6},
  year={2014}
}

@inproceedings{frameworks-highlevel-cusha,
  title={CuSha: vertex-centric graph processing on GPUs},
  author={Khorasani, Farzad and Vora, Keval and Gupta, Rajiv and Bhuyan, Laxmi N},
  booktitle={Proceedings of the 23rd international symposium on High-performance parallel and distributed computing},
  pages={239--252},
  year={2014}
}

@article{frameworks-highlevel-gunrock,
  title={Gunrock: GPU graph analytics},
  author={Wang, Yangzihao and Pan, Yuechao and Davidson, Andrew and Wu, Yuduo and Yang, Carl and Wang, Leyuan and Osama, Muhammad and Yuan, Chenshan and Liu, Weitang and Riffel, Andy T and others},
  journal={ACM Transactions on Parallel Computing (TOPC)},
  volume={4},
  number={1},
  pages={1--49},
  year={2017},
  publisher={ACM New York, NY, USA}
}

@inproceedings{frameworks-highlevel-compiler,
  title={A compiler for throughput optimization of graph algorithms on GPUs},
  author={Pai, Sreepathi and Pingali, Keshav},
  booktitle={Proceedings of the 2016 ACM SIGPLAN International Conference on Object-Oriented Programming, Systems, Languages, and Applications},
  pages={1--19},
  year={2016}
}

@inproceedings{frameworks-highlevel-gts,
  title={GTS: A fast and scalable graph processing method based on streaming topology to GPUs},
  author={Kim, Min-Soo and An, Kyuhyeon and Park, Himchan and Seo, Hyunseok and Kim, Jinwook},
  booktitle={Proceedings of the 2016 International Conference on Management of Data},
  pages={447--461},
  year={2016}
}

@inproceedings{frameworks-highlevel-yoke,
  title={A yoke of oxen and a thousand chickens for heavy lifting graph processing},
  author={Gharaibeh, Abdullah and Beltr{\~a}o Costa, Lauro and Santos-Neto, Elizeu and Ripeanu, Matei},
  booktitle={Proceedings of the 21st international conference on Parallel architectures and compilation techniques},
  pages={345--354},
  year={2012}
}

@article{frameworks-highlevel-gpudistributed,
  title={A distributed multi-gpu system for fast graph processing},
  author={Jia, Zhihao and Kwon, Yongkee and Shipman, Galen and McCormick, Pat and Erez, Mattan and Aiken, Alex},
  journal={Proceedings of the VLDB Endowment},
  volume={11},
  number={3},
  pages={297--310},
  year={2017},
  publisher={VLDB Endowment}
}

@article{frameworks-highlevel-groute,
  title={Groute: An asynchronous multi-GPU programming model for irregular computations},
  author={Ben-Nun, Tal and Sutton, Michael and Pai, Sreepathi and Pingali, Keshav},
  journal={ACM SIGPLAN Notices},
  volume={52},
  number={8},
  pages={235--248},
  year={2017},
  publisher={ACM New York, NY, USA}
}

@inproceedings{frameworks-highlevel-multigraph,
  title={Multigraph: Efficient graph processing on gpus},
  author={Hong, Changwan and Sukumaran-Rajam, Aravind and Kim, Jinsung and Sadayappan, P},
  booktitle={2017 26th International Conference on Parallel Architectures and Compilation Techniques (PACT)},
  pages={27--40},
  year={2017},
  organization={IEEE}
}

@inproceedings{frameworks-highlevel-digraph,
  title={DiGraph: An efficient path-based iterative directed graph processing system on multiple GPUs},
  author={Zhang, Yu and Liao, Xiaofei and Jin, Hai and He, Bingsheng and Liu, Haikun and Gu, Lin},
  booktitle={Proceedings of the Twenty-Fourth International Conference on Architectural Support for Programming Languages and Operating Systems},
  pages={601--614},
  year={2019}
}

@article{frameworks-highlevel-emptyheaded,
  title={EmptyHeaded: boolean algebra based graph processing},
  author={Aberger, Christopher R and N{\"o}tzli, Andres and Olukotun, Kunle and R{\'e}, Christopher},
  journal={ArXiv e-prints},
  year={2015}
}

@article{frameworks-highlevel-BGL,
  title={The parallel BGL: A generic library for distributed graph computations},
  author={Gregor, Douglas and Lumsdaine, Andrew},
  journal={Parallel Object-Oriented Scientific Computing (POOSC)},
  volume={2},
  pages={1--18},
  year={2005},
  publisher={Citeseer}
}

@article{frameworks-highlevel-gstream,
  title={Gstream: A graph streaming processing method for large-scale graphs on gpus},
  author={Seo, Hyunseok and Kim, Jinwook and Kim, Min-Soo},
  journal={ACM SIGPLAN Notices},
  volume={50},
  number={8},
  pages={253--254},
  year={2015},
  publisher={ACM New York, NY, USA}
}

@inproceedings{frameworks-highlevel-ligra,
  title={Ligra: a lightweight graph processing framework for shared memory},
  author={Shun, Julian and Blelloch, Guy E},
  booktitle={Proceedings of the 18th ACM SIGPLAN symposium on Principles and practice of parallel programming},
  pages={135--146},
  year={2013}
}

@inproceedings{frameworks-highlevel-efficientcpu-gpu,
  title={Efficient parallel graph exploration on multi-core CPU and GPU},
  author={Hong, Sungpack and Oguntebi, Tayo and Olukotun, Kunle},
  booktitle={2011 International Conference on Parallel Architectures and Compilation Techniques},
  pages={78--88},
  year={2011},
  organization={IEEE}
}

@inproceedings{frameworks-highlevel-gbtl,
  title={GBTL-CUDA: Graph algorithms and primitives for GPUs},
  author={Zhang, Peter and Zalewski, Marcin and Lumsdaine, Andrew and Misurda, Samantha and McMillan, Scott},
  booktitle={2016 IEEE International Parallel and Distributed Processing Symposium Workshops (IPDPSW)},
  pages={912--920},
  year={2016},
  organization={IEEE}
}

@inproceedings{frameworks-highlevel-towards,
  title={Towards GPU-accelerated large-scale graph processing in the cloud},
  author={Zhong, Jianlong and He, Bingsheng},
  booktitle={2013 IEEE 5th International Conference on Cloud Computing Technology and Science},
  volume={1},
  pages={9--16},
  year={2013},
  organization={IEEE}
}

@inproceedings{frameworks-highlevel-scalablesimd,
  title={Scalable simd-efficient graph processing on gpus},
  author={Khorasani, Farzad and Gupta, Rajiv and Bhuyan, Laxmi N},
  booktitle={2015 International Conference on Parallel Architecture and Compilation (PACT)},
  pages={39--50},
  year={2015},
  organization={IEEE}
}

@inproceedings{sparse-clspmv,
  title={clSpMV: A cross-platform OpenCL SpMV framework on GPUs},
  author={Su, Bor-Yiing and Keutzer, Kurt},
  booktitle={Proceedings of the 26th ACM international conference on Supercomputing},
  pages={353--364},
  year={2012}
}

@inproceedings{sparse-tensorcore,
  title={Sparse tensor core: Algorithm and hardware co-design for vector-wise sparse neural networks on modern gpus},
  author={Zhu, Maohua and Zhang, Tao and Gu, Zhenyu and Xie, Yuan},
  booktitle={Proceedings of the 52nd Annual IEEE/ACM International Symposium on Microarchitecture},
  pages={359--371},
  year={2019}
}

@article{sparse-framework-yaSpMV,
  title={yaSpMV: yet another SpMV framework on GPUs},
  author={Yan, Shengen and Li, Chao and Zhang, Yunquan and Zhou, Huiyang},
  journal={Acm Sigplan Notices},
  volume={49},
  number={8},
  pages={107--118},
  year={2014},
  publisher={ACM New York, NY, USA}
}

@article{sparse-framework-biell,
  title={BiELL: A bisection ELLPACK-based storage format for optimizing SpMV on GPUs},
  author={Zheng, Cong and Gu, Shuo and Gu, Tong-Xiang and Yang, Bing and Liu, Xing-Ping},
  journal={Journal of Parallel and Distributed Computing},
  volume={74},
  number={7},
  pages={2639--2647},
  year={2014},
  publisher={Elsevier}
}

@inproceedings{sparse-framework-fastsparse,
  title={Fast sparse matrix-vector multiplication on GPUs for graph applications},
  author={Ashari, Arash and Sedaghati, Naser and Eisenlohr, John and Parthasarath, Srinivasan and Sadayappan, P},
  booktitle={SC'14: Proceedings of the International Conference for High Performance Computing, Networking, Storage and Analysis},
  pages={781--792},
  year={2014},
  organization={IEEE}
}

@inproceedings{sparse-framework-implementingsparse,
  title={Implementing sparse matrix-vector multiplication on throughput-oriented processors},
  author={Bell, Nathan and Garland, Michael},
  booktitle={Proceedings of the conference on high performance computing networking, storage and analysis},
  pages={1--11},
  year={2009}
}

@article{sparse-framework-sparsematrix,
  title={Sparse matrix-vector multiplication on GPGPUs},
  author={Filippone, Salvatore and Cardellini, Valeria and Barbieri, Davide and Fanfarillo, Alessandro},
  journal={ACM Transactions on Mathematical Software (TOMS)},
  volume={43},
  number={4},
  pages={1--49},
  year={2017},
  publisher={ACM New York, NY, USA}
}

@article{sparse-framework-performance,
  title={Performance analysis and optimization for SpMV on GPU using probabilistic modeling},
  author={Li, Kenli and Yang, Wangdong and Li, Keqin},
  journal={IEEE Transactions on Parallel and Distributed Systems},
  volume={26},
  number={1},
  pages={196--205},
  year={2014},
  publisher={IEEE}
}

@misc{sparse-lib-mkl,
  title = {MKL sparse-BLAS library},
  author ={Intel},
  howpublished = {\url{http://software.intel.com/en-us/intel-mkl}},
  year={2020}
}

@inproceedings{accelerator-pim-scalable,
  title={A scalable processing-in-memory accelerator for parallel graph processing},
  author={Ahn, Junwhan and Hong, Sungpack and Yoo, Sungjoo and Mutlu, Onur and Choi, Kiyoung},
  booktitle={Proceedings of the 42nd Annual International Symposium on Computer Architecture},
  pages={105--117},
  year={2015}
}

@inproceedings{accelerator-pim-graphp,
  title={GraphP: Reducing communication for PIM-based graph processing with efficient data partition},
  author={Zhang, Mingxing and Zhuo, Youwei and Wang, Chao and Gao, Mingyu and Wu, Yongwei and Chen, Kang and Kozyrakis, Christos and Qian, Xuehai},
  booktitle={2018 IEEE International Symposium on High Performance Computer Architecture (HPCA)},
  pages={544--557},
  year={2018},
  organization={IEEE}
}

@article{accelerator-pim-graphh,
  title={Graphh: A processing-in-memory architecture for large-scale graph processing},
  author={Dai, Guohao and Huang, Tianhao and Chi, Yuze and Zhao, Jishen and Sun, Guangyu and Liu, Yongpan and Wang, Yu and Xie, Yuan and Yang, Huazhong},
  journal={IEEE Transactions on Computer-Aided Design of Integrated Circuits and Systems},
  volume={38},
  number={4},
  pages={640--653},
  year={2018},
  publisher={IEEE}
}

@inproceedings{accelerator-algebra-efficient,
  title={Efficient pagerank and spmv computation on amd gpus},
  author={Wu, Tianji and Wang, Bo and Shan, Yi and Yan, Feng and Wang, Yu and Xu, Ningyi},
  booktitle={2010 39th International Conference on Parallel Processing},
  pages={81--89},
  year={2010},
  organization={IEEE}
}

@inproceedings{accelerator-algebra-slimsell,
  title={Slimsell: A vectorizable graph representation for breadth-first search},
  author={Besta, Maciej and Marending, Florian and Solomonik, Edgar and Hoefler, Torsten},
  booktitle={2017 IEEE International Parallel and Distributed Processing Symposium (IPDPS)},
  pages={32--41},
  year={2017},
  organization={IEEE}
}

@inproceedings{accelerator-pim-graphpim,
  title={Graphpim: Enabling instruction-level pim offloading in graph computing frameworks},
  author={Nai, Lifeng and Hadidi, Ramyad and Sim, Jaewoong and Kim, Hyojong and Kumar, Pranith and Kim, Hyesoon},
  booktitle={2017 IEEE International symposium on high performance computer architecture (HPCA)},
  pages={457--468},
  year={2017},
  organization={IEEE}
}

@inproceedings{accelerator-pim-sisa,
  title={Sisa: Set-centric instruction set architecture for graph mining on processing-in-memory systems},
  author={Besta, Maciej and Kanakagiri, Raghavendra and Kwasniewski, Grzegorz and Ausavarungnirun, Rachata and Ber{\'a}nek, Jakub and Kanellopoulos, Konstantinos and Janda, Kacper and Vonarburg-Shmaria, Zur and Gianinazzi, Lukas and Stefan, Ioana and others},
  booktitle={MICRO-54: 54th Annual IEEE/ACM International Symposium on Microarchitecture},
  pages={282--297},
  year={2021}
}

@article{accelerator-pim-pim-enabled,
  title={PIM-enabled instructions: A low-overhead, locality-aware processing-in-memory architecture},
  author={Ahn, Junwhan and Yoo, Sungjoo and Mutlu, Onur and Choi, Kiyoung},
  journal={ACM SIGARCH Computer Architecture News},
  volume={43},
  number={3S},
  pages={336--348},
  year={2015},
  publisher={ACM New York, NY, USA}
}

@inproceedings{accelerator-pim-conda,
  title={CoNDA: Efficient cache coherence support for near-data accelerators},
  author={Boroumand, Amirali and Ghose, Saugata and Patel, Minesh and Hassan, Hasan and Lucia, Brandon and Ausavarungnirun, Rachata and Hsieh, Kevin and Hajinazar, Nastaran and Malladi, Krishna T and Zheng, Hongzhong and others},
  booktitle={Proceedings of the 46th International Symposium on Computer Architecture},
  pages={629--642},
  year={2019}
}

@article{accelerator-pim-lazy,
  title={LazyPIM: An efficient cache coherence mechanism for processing-in-memory},
  author={Boroumand, Amirali and Ghose, Saugata and Patel, Minesh and Hassan, Hasan and Lucia, Brandon and Hsieh, Kevin and Malladi, Krishna T and Zheng, Hongzhong and Mutlu, Onur},
  journal={IEEE Computer Architecture Letters},
  volume={16},
  number={1},
  pages={46--50},
  year={2016},
  publisher={IEEE}
}

@inproceedings{accelerator-pim-graphq,
  title={Graphq: Scalable pim-based graph processing},
  author={Zhuo, Youwei and Wang, Chao and Zhang, Mingxing and Wang, Rui and Niu, Dimin and Wang, Yanzhi and Qian, Xuehai},
  booktitle={Proceedings of the 52nd Annual IEEE/ACM International Symposium on Microarchitecture},
  pages={712--725},
  year={2019}
}

@inproceedings{sparse-pim-spaceA,
  title={SpaceA: Sparse matrix vector multiplication on processing-in-memory accelerator},
  author={Xie, Xinfeng and Liang, Zheng and Gu, Peng and Basak, Abanti and Deng, Lei and Liang, Ling and Hu, Xing and Xie, Yuan},
  booktitle={2021 IEEE International Symposium on High-Performance Computer Architecture (HPCA)},
  pages={570--583},
  year={2021},
  organization={IEEE}
}

@inproceedings{sparse-pim-Abc,
  title={Abc-dimm: Alleviating the bottleneck of communication in dimm-based near-memory processing with inter-dimm broadcast},
  author={Sun, Weiyi and Li, Zhaoshi and Yin, Shouyi and Wei, Shaojun and Liu, Leibo},
  booktitle={2021 ACM/IEEE 48th Annual International Symposium on Computer Architecture (ISCA)},
  pages={237--250},
  year={2021},
  organization={IEEE}
}

@inproceedings{sparse-pim-3d-stacked,
  title={Accelerating sparse matrix-matrix multiplication with 3D-stacked logic-in-memory hardware},
  author={Zhu, Qiuling and Graf, Tobias and Sumbul, H Ekin and Pileggi, Larry and Franchetti, Franz},
  booktitle={2013 IEEE High Performance Extreme Computing Conference (HPEC)},
  pages={1--6},
  year={2013},
  organization={IEEE}
}

@inproceedings{sparse-pim-gpu,
  title={Efficient sparse-matrix multi-vector product on gpus},
  author={Hong, Changwan and Sukumaran-Rajam, Aravind and Bandyopadhyay, Bortik and Kim, Jinsung and Kurt, S{\"u}reyya Emre and Nisa, Israt and Sabhlok, Shivani and {\c{C}}ataly{\"u}rek, {\"U}mit V and Parthasarathy, Srinivasan and Sadayappan, P},
  booktitle={Proceedings of the 27th International Symposium on High-Performance Parallel and Distributed Computing},
  pages={66--79},
  year={2018}
}

@article{related-framework-graphblast,
  title={GraphBLAST: A high-performance linear algebra-based graph framework on the GPU},
  author={Yang, Carl and Buluc, Aydin and Owens, John D},
  journal={arXiv preprint arXiv:1908.01407},
  year={2019}
}

@article{related-pimpam,
  title={PimPam: Efficient Graph Pattern Matching on Real Processing-in-Memory Hardware},
  author={Cai, Shuangyu and Tian, Boyu and Zhang, Huanchen and Gao, Mingyu},
  journal={Proceedings of the ACM on Management of Data},
  volume={2},
  number={3},
  pages={1--25},
  year={2024},
  publisher={ACM New York, NY, USA}
}

@inproceedings{related-gaas,
  title={GaaS-X: Graph analytics accelerator supporting sparse data representation using crossbar architectures},
  author={Challapalle, Nagadastagiri and Rampalli, Sahithi and Song, Linghao and Chandramoorthy, Nandhini and Swaminathan, Karthik and Sampson, John and Chen, Yiran and Narayanan, Vijaykrishnan},
  booktitle={2020 ACM/IEEE 47th Annual International Symposium on Computer Architecture (ISCA)},
  pages={433--445},
  year={2020},
  organization={IEEE}
}

@inproceedings{related-psyncpim,
  title={pSyncPIM: Partially Synchronous Execution of Sparse Matrix Operations for All-Bank PIM Architectures},
  author={Baek, Daehyeon and Hwang, Soojin and Huh, Jaehyuk},
  booktitle={2024 ACM/IEEE 51st Annual International Symposium on Computer Architecture (ISCA)},
  pages={354--367},
  year={2024},
  organization={IEEE}
}

@article{intro-graphapp-social-polites2008centrality,
  title={The centrality and prestige of CACM},
  author={Polites, Greta L and Watson, Richard T},
  journal={Communications of the ACM},
  volume={51},
  number={1},
  pages={95--100},
  year={2008},
  publisher={ACM New York, NY, USA}
}

@inproceedings{intro-graphapp-social-azad2015parallel,
  title={A parallel tree grafting algorithm for maximum cardinality matching in bipartite graphs},
  author={Azad, Ariful and Bulu{\c{c}}, Aydin and Pothen, Alex},
  booktitle={2015 IEEE International Parallel and Distributed Processing Symposium},
  pages={1075--1084},
  year={2015},
  organization={IEEE}
}

@inproceedings{intro-graphapp-NLP-manning2014stanford,
  title={The Stanford CoreNLP natural language processing toolkit},
  author={Manning, Christopher D and Surdeanu, Mihai and Bauer, John and Finkel, Jenny Rose and Bethard, Steven and McClosky, David},
  booktitle={Proceedings of 52nd annual meeting of the association for computational linguistics: system demonstrations},
  pages={55--60},
  year={2014}
}

@inproceedings{intro-graphapp-NLP-hahn1984computing,
  title={Computing text constituency: An algorithmic approach to the generation of text graphs},
  author={Hahn, Udo and Reimer, Ulrich},
  booktitle={Proceedings of the 7th annual international ACM SIGIR conference on Research and development in information retrieval},
  pages={343--368},
  year={1984}
}

@article{intro-graphapp-NLP-miller1995wordnet,
  title={WordNet: a lexical database for English},
  author={Miller, George A},
  journal={Communications of the ACM},
  volume={38},
  number={11},
  pages={39--41},
  year={1995},
  publisher={ACM New York, NY, USA}
}

@article{intro-graphapp-chemical-ivanciuc1999graph,
  title={The graph description of chemical structures},
  author={Ivanciuc, O and Balaban, AT},
  journal={Topological indices and related descriptors in QSAR and QSPR},
  pages={59--167},
  year={1999}
}

@book{background-linearAlgebraicGraph,
  title={Graph algorithms in the language of linear algebra},
  author={Kepner, Jeremy and Gilbert, John},
  year={2011},
  publisher={SIAM}
}

@book{background-semiring1,
  address = {Reading, Mass.},
  author = {Aho, Alfred V. and Hopcroft, John E. and Ullman, Jeffrey D.},
  publisher = {Addison-Wesley},
  title = {{The Design and Analysis of Computer Algorithms}},
  year = 1974
}

@book{background-semiring2,
  title={Introduction to algorithms},
  author={Thomas H.. Cormen and Leiserson, Charles Eric and Rivest, Ronald L and Stein, Clifford},
  volume={5},
  year={2001},
  publisher={MIT press Cambridge}
}

@book{background-semiring3,
  title={Graphs, dioids and semirings: new models and algorithms},
  author={Gondran, Michel and Minoux, Michel},
  volume={41},
  year={2008},
  publisher={Springer Science \& Business Media}
}

@article{background-semiring4,
  title={Dio{\"\i}ds and semirings: Links to fuzzy sets and other applications},
  author={Gondran, Michel and Minoux, Michel},
  journal={Fuzzy Sets and Systems},
  volume={158},
  number={12},
  pages={1273--1294},
  year={2007},
  publisher={Elsevier}
}

@inproceedings{background-semiring5,
  title={Fun with semirings: a functional pearl on the abuse of linear algebra},
  author={Dolan, Stephen},
  booktitle={Proceedings of the 18th ACM SIGPLAN international conference on Functional programming},
  pages={101--110},
  year={2013}
}

@article{background-semiring6,
  title={Commuting graphs of matrices over semirings},
  author={Dol{\v{z}}an, David and Oblak, Polona},
  journal={Linear algebra and its applications},
  volume={435},
  number={7},
  pages={1657--1665},
  year={2011},
  publisher={Elsevier}
}

@inproceedings{background-upmem,
  title={Introduction to UPMEM PIM. Processing-in-memory (PIM) on DRAM Accelerator (White Paper)},
  author={UPMEM},
  booktitle={},
  year={2018}
}

@misc{background-upmem-website,
  title ={UPMEM Website},
  author={UPMEM},
  howpublished = {\url{https:// www.upmem.com}},
  year={2020}
}

@inproceedings{background-csr,
  title={Sparse matrix format selection with multiclass SVM for SpMV on GPU},
  author={Benatia, Akrem and Ji, Weixing and Wang, Yizhuo and Shi, Feng},
  booktitle={2016 45th International Conference on Parallel Processing (ICPP)},
  pages={496--505},
  year={2016},
  organization={IEEE}
}

@article{background-coo-csr,
  title={A survey of indexing techniques for sparse matrices},
  author={Pooch, Udo W and Nieder, Al},
  journal={ACM Computing Surveys (CSUR)},
  volume={5},
  number={2},
  pages={109--133},
  year={1973},
  publisher={ACM New York, NY, USA}
}

@article{background-coo,
  title={Scan primitives for GPU computing},
  author={Sengupta, Shubhabrata and Harris, Mark and Zhang, Yao and Owens, John D},
  year={2007}
}

@inproceedings{app-bfs-local,
  title={A local search mechanism for peer-to-peer networks},
  author={Kalogeraki, Vana and Gunopulos, Dimitrios and Zeinalipour-Yazti, Demetrios},
  booktitle={Proceedings of the eleventh international conference on Information and knowledge management},
  pages={300--307},
  year={2002}
}

@inproceedings{app-bfs-improving,
  title={Improving search in peer-to-peer networks},
  author={Yang, Beverly and Garcia-Molina, Hector},
  booktitle={Proceedings 22nd International Conference on Distributed Computing Systems},
  pages={5--14},
  year={2002},
  organization={IEEE}
}

@misc{app-sssp-fast,
  title={Fast Route Planning},
  author={Google Tech Talk },
  year={2009}
}

@inproceedings{app-sssp-highway,
  title={Highway dimension, shortest paths, and provably efficient algorithms},
  author={Abraham, Ittai and Fiat, Amos and Goldberg, Andrew V and Werneck, Renato F},
  booktitle={Proceedings of the twenty-first annual ACM-SIAM symposium on Discrete Algorithms},
  pages={782--793},
  year={2010},
  organization={SIAM}
}

@inproceedings{app-sssp-hub,
  title={A hub-based labeling algorithm for shortest paths in road networks},
  author={Abraham, Ittai and Delling, Daniel and Goldberg, Andrew V and Werneck, Renato F},
  booktitle={International Symposium on Experimental Algorithms},
  pages={230--241},
  year={2011},
  organization={Springer}
}

@techreport{app-pr-pagerank,
  title={The PageRank citation ranking: Bringing order to the web.},
  author={Page, Lawrence and Brin, Sergey and Motwani, Rajeev and Winograd, Terry},
  year={1999},
  institution={Stanford InfoLab}
}

@misc{app-nvgraph,
  title = {nvGRAPH Library},
  author={NVIDIA},
  howpublished = {\url{https://developer.nvidia.com/nvgraph}},
  year={2019}
}

@misc{cusparse,
  title = {NVIDIA's cuSPARSE},
  author={NVIDIA},
  howpublished = {\url{https://docs.nvidia.com/cuda/cusparse/index.html}},
  year={2020}
}

@misc{app-cugraph,
  title = {cuGraph Library},
  author={RAPIDS},
  howpublished = {\url{https://github.com/rapidsai/nvgraph}},
  year={2021}
}

@misc{app-RAPIDS,
  title = {RAPIDS},
  author={DataScienceFramework },
  howpublished = {\url{https://rapids.ai/start.html}},
  year={2021}
}

@incollection{app-basic-graph,
  title={Graph algorithms},
  author={van Leeuwen, Jan},
  booktitle={Algorithms and complexity},
  pages={525--631},
  year={1990},
  publisher={Elsevier}
}

@article{app-basic-graphnetwork,
  title={Graph and network algorithms},
  author={Khuller, Samir and Raghavachari, Balaji},
  journal={ACM Computing Surveys (CSUR)},
  volume={28},
  number={1},
  pages={43--45},
  year={1996},
  publisher={ACM New York, NY, USA}
}

@article{app-ppr,
  title={Efficient Algorithms for Personalized PageRank Computation: A Survey},
  author={Yang, Mingji and Wang, Hanzhi and Wei, Zhewei and Wang, Sibo and Wen, Ji-Rong},
  journal={IEEE Transactions on Knowledge and Data Engineering},
  year={2024},
  publisher={IEEE}
}

@inproceedings{tools-ddr4,
  title={JESD79-4 DDR4 SDRAM standard)},
  author={JEDEC},
  booktitle={JEDEC},
  year={2012}
}

@misc{tools-cpu,
  title ={Intel Xeon Silver 4110 Processor},
  author={Intel},
  howpublished = {\url{https:// ark.intel.com/ content/ www/ us/ en/ ark/ products/ 123547/ intel-xeon-silver-
4110-processor-11m-cache-2-10-ghz.html}},
  year={2017}
}

@misc{tools-cpu-gridgraph,
  title ={Intel® Core™ i7-1265U},
  author={Intel},
  howpublished = {\url{https://www.intel.com/content/www/us/en/products/sku/226258/intel-core-i71265u-processor-12m-cache-up-to-4-80-ghz/specifications.html}},
  year={2022}
}

@misc{tools-cpu-power,
  title ={RAPL Power Meter},
  author={Intel Open Source},
  howpublished = {\url{https://www.intel.com/content/www/us/en/developer/topic-technology/open/overview.html}},
  year={2014}
}

@misc{tools-gpu-power,
  title ={NVIDIA System Management Interface Program},
  author={Intel Open Source},
  howpublished = {\url{https://developer.download.nvidia.com/compute/DCGM/docs/nvidia-smi-367.38.pdf}},
  year={2016}
}

@article{tools-pim-power,
  title={Intel{\textregistered} 64 and ia-32 architectures software developer’s manual},
  author={Guide, Part},
  journal={Volume 3B: System programming Guide, Part},
  volume={2},
  number={11},
  pages={0--40},
  year={2011}
}

@misc{tools-gpu,
  title ={NVIDIA RTX 3050},
  author={NVIDIA},
  howpublished = {\url{https://www.nvidia.com/en-us/geforce/graphics-cards/30-series/rtx-3050/}},
  year={2022}
}

@inproceedings{tools-upimulator,
  title={Pathfinding Future PIM Architectures by Demystifying a Commercial PIM Technology},
  author={Hyun, Bongjoon and Kim, Taehun and Lee, Dongjae and Rhu, Minsoo},
  booktitle={2024 IEEE International Symposium on High-Performance Computer Architecture (HPCA)},
  pages={263--279},
  year={2024},
  organization={IEEE}
}

@misc{tools-peakperf,
  title ={PeakPerf},
  author={PeakPerf},
  howpublished = {\url{https://github.com/Dr-Noob/peakperf}},
  year={2021}
}

@inproceedings{compare-cpu-gridgraph,
  title={$\{$GridGraph$\}$:$\{$Large-Scale$\}$ graph processing on a single machine using 2-level hierarchical partitioning},
  author={Zhu, Xiaowei and Han, Wentao and Chen, Wenguang},
  booktitle={2015 USENIX Annual Technical Conference (USENIX ATC 15)},
  pages={375--386},
  year={2015}
}

@article{pim-sparsep,
  title={Sparsep: Towards efficient sparse matrix vector multiplication on real processing-in-memory architectures},
  author={Giannoula, Christina and Fernandez, Ivan and Luna, Juan G{\'o}mez and Koziris, Nectarios and Goumas, Georgios and Mutlu, Onur},
  journal={Proceedings of the ACM on Measurement and Analysis of Computing Systems},
  volume={6},
  number={1},
  pages={1--49},
  year={2022},
  publisher={ACM New York, NY, USA}
}

@inproceedings{spmspv-parallel,
  title={A work-efficient parallel sparse matrix-sparse vector multiplication algorithm},
  author={Azad, Ariful and Bulu{\c{c}}, Aydin},
  booktitle={2017 IEEE International Parallel and Distributed Processing Symposium (IPDPS)},
  pages={688--697},
  year={2017},
  organization={IEEE}
}

@inproceedings{spmspv-tile,
  title={Tilespmspv: A tiled algorithm for sparse matrix-sparse vector multiplication on gpus},
  author={Ji, Haonan and Song, Huimin and Lu, Shibo and Jin, Zhou and Tan, Guangming and Liu, Weifeng},
  booktitle={Proceedings of the 51st International Conference on Parallel Processing},
  pages={1--11},
  year={2022}
}

@misc{datasets-graph,
  title ={GraphChallenge Website},
  author={GraphChallenge},
  howpublished = {\url{https://graphchallenge.mit.edu/data-sets}},
  year={2003}
}

@misc{datasets-graph-snap,
  title ={Stanford Large Network Dataset Collection Website},
  author={Stanford Large Network Dataset Collection},
  howpublished = {\url{https://snap.stanford.edu/data/}},
  year={2009}
}

@inproceedings{motivation-bulk-anavaram,
  title={Graph processing on GPUs: Where are the bottlenecks?},
  author={Xu, Qiumin and Jeon, Hyeran and Annavaram, Murali},
  booktitle={2014 IEEE International Symposium on Workload Characterization (IISWC)},
  pages={140--149},
  year={2014},
  organization={IEEE}
}

@inproceedings{motivaion-spmspv-parallel,
  title={Parallel breadth-first search on distributed memory systems},
  author={Bulu{\c{c}}, Aydin and Madduri, Kamesh},
  booktitle={Proceedings of 2011 International Conference for High Performance Computing, Networking, Storage and Analysis},
  pages={1--12},
  year={2011}
}

@inproceedings{motivation-spmspv-owen,
  title={Implementing push-pull efficiently in GraphBLAS},
  author={Yang, Carl and Bulu{\c{c}}, Ayd{\i}n and Owens, John D},
  booktitle={Proceedings of the 47th International Conference on Parallel Processing},
  pages={1--11},
  year={2018}
}

@inproceedings{motivation-spmspv-fast,
  title={Fast sparse matrix and sparse vector multiplication algorithm on the GPU},
  author={Yang, Carl and Wang, Yangzihao and Owens, John D},
  booktitle={2015 IEEE International Parallel and Distributed Processing Symposium Workshop},
  pages={841--847},
  year={2015},
  organization={IEEE}
}

@inproceedings{motivaiton-pim-co-ml,
  title={Co-ML: a case for Co llaborative ML acceleration using near-data processing},
  author={Aga, Shaizeen and Jayasena, Nuwan and Ignatowski, Mike},
  booktitle={Proceedings of the International Symposium on Memory Systems},
  pages={506--517},
  year={2019}
}

@inproceedings{motivation-pim-google,
  title={Google workloads for consumer devices: Mitigating data movement bottlenecks},
  author={Boroumand, Amirali and Ghose, Saugata and Kim, Youngsok and Ausavarungnirun, Rachata and Shiu, Eric and Thakur, Rahul and Kim, Daehyun and Kuusela, Aki and Knies, Allan and Ranganathan, Parthasarathy and others},
  booktitle={Proceedings of the Twenty-Third International Conference on Architectural Support for Programming Languages and Operating Systems},
  pages={316--331},
  year={2018}
}

@inproceedings{motivation-pim-concurrent,
  title={Concurrent data structures with near-data-processing: An architecture-aware implementation},
  author={Choe, Jiwon and Huang, Amy and Moreshet, Tali and Herlihy, Maurice and Bahar, R Iris},
  booktitle={The 31st ACM Symposium on Parallelism in Algorithms and Architectures},
  pages={297--308},
  year={2019}
}

@inproceedings{motivation-pim-neural,
  title={Neural cache: Bit-serial in-cache acceleration of deep neural networks},
  author={Eckert, Charles and Wang, Xiaowei and Wang, Jingcheng and Subramaniyan, Arun and Iyer, Ravi and Sylvester, Dennis and Blaaauw, David and Das, Reetuparna},
  booktitle={2018 ACM/IEEE 45Th annual international symposium on computer architecture (ISCA)},
  pages={383--396},
  year={2018},
  organization={IEEE}
}

@inproceedings{motivation-pim-nda,
  title={NDA: Near-DRAM acceleration architecture leveraging commodity DRAM devices and standard memory modules},
  author={Farmahini-Farahani, Amin and Ahn, Jung Ho and Morrow, Katherine and Kim, Nam Sung},
  booktitle={2015 IEEE 21st International Symposium on High Performance Computer Architecture (HPCA)},
  pages={283--295},
  year={2015},
  organization={IEEE}
}

@article{motivation-pim-processing,
  title={Processing-in-memory: A workload-driven perspective},
  author={Ghose, Saugata and Boroumand, Amirali and Kim, Jeremie S and G{\'o}mez-Luna, Juan and Mutlu, Onur},
  journal={IBM Journal of Research and Development},
  volume={63},
  number={6},
  pages={3--1},
  year={2019},
  publisher={IBM}
}

@inproceedings{motivation-pim-continuous,
  title={Continuous runahead: Transparent hardware acceleration for memory intensive workloads},
  author={Hashemi, Milad and Mutlu, Onur and Patt, Yale N},
  booktitle={2016 49th Annual IEEE/ACM International Symposium on Microarchitecture (MICRO)},
  pages={1--12},
  year={2016},
  organization={IEEE}
}

@inproceedings{ipc-memory-bound-comparison,
  title={The comparison of large-scale graph processing algorithms implementation methods for Intel KNL and NVIDIA GPU},
  author={Afanasyev, Ilya and Voevodin, Vladimir},
  booktitle={Russian Supercomputing Days},
  pages={80--94},
  year={2017},
  organization={Springer}
}

@inproceedings{ipc-memory-bound-energy,
  title={The energy case for graph processing on hybrid cpu and gpu systems},
  author={Gharaibeh, Abdullah and Santos-Neto, Elizeu and Costa, Lauro Beltr{\~a}o and Ripeanu, Matei},
  booktitle={Proceedings of the 3rd Workshop on Irregular Applications: Architectures and Algorithms},
  pages={1--8},
  year={2013}
}

@article{memory-bound-sparse,
  title={Sparse matrix libraries in C++ for high performance architectures},
  author={Dongarraxz, Jack and Lumsdaine, Andrew and Niu, Xinhui and Pozoz, Roldan and Remingtonx, Karin},
  year={1994}
}

@inproceedings{memory-bound-multicore,
  title={Performance analysis and optimization of sparse matrix-vector multiplication on modern multi-and many-core processors},
  author={Elafrou, Athena and Goumas, Georgios and Koziris, Nectarios},
  booktitle={2017 46th International Conference on Parallel Processing (ICPP)},
  pages={292--301},
  year={2017},
  organization={IEEE}
}

@inproceedings{memory-bound-conflict,
  title={Conflict-free symmetric sparse matrix-vector multiplication on multicore architectures},
  author={Elafrou, Athena and Goumas, Georgios and Koziris, Nectarios},
  booktitle={Proceedings of the International Conference for High Performance Computing, Networking, Storage and Analysis},
  pages={1--15},
  year={2019}
}

@ARTICLE{memory-bound-benchmarking,
  author={Gómez-Luna, Juan and Hajj, Izzat El and Fernandez, Ivan and Giannoula, Christina and Oliveira, Geraldo F. and Mutlu, Onur},
  journal={IEEE Access}, 
  title={Benchmarking a New Paradigm: Experimental Analysis and Characterization of a Real Processing-in-Memory System}, 
  year={2022},
  volume={10},
  number={},
  pages={52565-52608},
  doi={10.1109/ACCESS.2022.3174101}}

@inproceedings{memory-bound-benchmarking2,
  title={Benchmarking memory-centric computing systems: Analysis of real processing-in-memory hardware},
  author={G{\'o}mez-Luna, Juan and El Hajj, Izzat and Fernandez, Ivan and Giannoula, Christina and Oliveira, Geraldo F and Mutlu, Onur},
  booktitle={2021 12th International Green and Sustainable Computing Conference (IGSC)},
  pages={1--7},
  year={2021},
  organization={IEEE}
}

@article{memory-bound-sparsity,
  title={Sparsity: Optimization framework for sparse matrix kernels},
  author={Im, Eun-Jin and Yelick, Katherine and Vuduc, Richard},
  journal={The International Journal of High Performance Computing Applications},
  volume={18},
  number={1},
  pages={135--158},
  year={2004},
  publisher={SAGE Publications}
}

@INPROCEEDINGS{SynCronHPCA2021,
  author={Giannoula, Christina and Vijaykumar, Nandita and Papadopoulou, Nikela and Karakostas, Vasileios and Fernandez, Ivan and Gómez-Luna, Juan and Orosa, Lois and Koziris, Nectarios and Goumas, Georgios and Mutlu, Onur},
  booktitle={2021 IEEE International Symposium on High-Performance Computer Architecture (HPCA)}, 
  title={SynCron: Efficient Synchronization Support for Near-Data-Processing Architectures}, 
  year={2021}
}

@article{PyGimSIGMETRICS25,
author = {Giannoula, Christina and Yang, Peiming and Fernandez, Ivan and Yang, Jiacheng and Durvasula, Sankeerth and Li, Yu Xin and Sadrosadati, Mohammad and Luna, Juan Gomez and Mutlu, Onur and Pekhimenko, Gennady},
title = {PyGim: An Efficient Graph Neural Network Library for Real Processing-In-Memory Architectures},
year = {2024},
volume = {8},
number = {3},
journal = {Proc. ACM Meas. Anal. Comput. Syst.},
month = dec,
articleno = {43},
}

\end{document}